\documentclass[superscriptaddress,prd,nofootinbib,preprintnumbers,longbibliography,11pt,eqsecnum]{revtex4-1}

\usepackage{amsmath} 
\usepackage{graphicx} 
\usepackage{amsthm}
\usepackage{amssymb} 
\usepackage{dsfont}
\usepackage{yfonts}
\usepackage{hyperref}
\usepackage{array,xcolor,graphicx}
\usepackage{booktabs,multirow}
\usepackage[utf8]{inputenc}
\usepackage{mathtools}

\usepackage{slashed}
\usepackage{caption}

\usepackage{etoolbox}
\patchcmd{\section}
  {\centering}
  {\raggedright}
  {}
  {}
\patchcmd{\subsection}
  {\centering}
  {\raggedright}
  {}
  {}

\usepackage[normalem]{ulem}
\usepackage{mathtools}

\usepackage[all]{xy}
\usepackage{tikz}
\usetikzlibrary{arrows.meta}

\hypersetup{colorlinks=true,linkcolor=blue,citecolor=red,urlcolor=blue, linktocpage=true}

\def\la{\langle}
\def\ra{\rangle}

\def\d{\partial}

\def\CE{\mathcal{E}}

\def\CN{\mathcal{N}}

\def\CT{\mathcal{T}}
\def\CO{\mathcal{O}}
\def\CP{\mathcal{P}}

\newcommand{\bln}{\begin{align}}
\newcommand{\eln}{\end{align}}
\newcommand{\bst}{\begin{split}}
\newcommand{\est}{\end{split}}
\newcommand{\bi}{\begin{itemize}}
\newcommand{\ei}{\end{itemize}}
\newcommand{\ben}{\begin{enumerate}}
\newcommand{\een}{\end{enumerate}}

\def\det{{\rm det}}

\def\eeq{\end{equation}}

\begin{document}

\title{First order non-Lorentzian fluids, entropy production and linear instabilities}

\author{Napat Poovuttikul}
\email{nickpoovuttikul@hi.is}
\author{Watse Sybesma}
\email{watse@hi.is}
\affiliation{University of Iceland, Science Institute, Dunhaga 3, IS-107, Reykjavik, Iceland}

\begin{abstract}
In this note, we investigate linear instabilities of hydrodynamics with corrections up to first order in derivatives. It has long been known that relativistic (Lorentzian) first order hydrodynamics, with positive local entropy production, exhibits unphysical instabilities. We extend this analysis to fluids with Galilean and Carrollian boost symmetries. We find that the instabilities occur in all cases, except for fluids with Galilean boost symmetry combined with the choice of macroscopic variables called Eckart frame. We also present a complete linearised analysis of the full spectrum of first order Carrollian hydrodynamics. Furthermore, we show that even in a fluid without boost symmetry present, instabilities can occur. These results provide evidence that the unphysical instabilities are symptoms of first order hydrodynamics, rather than a special feature of Lorentzian fluids.

\end{abstract}

\maketitle

\newpage
\begingroup
\tableofcontents
\endgroup

\section{Introduction}

Hydrodynamics is one of the most successful effective descriptions of many-body systems we have to date. Its applicability ranges from the scales of the galaxy, to the Earth's atmosphere, or to a glass of water, to a sample of a few millimetres of clean metal and graphene \cite{Moll16,Crossno1058,Bandurin1055}, to cold atomic systems (see, e.g., \cite{Adams:2012th} and reference therein), and even to subatomic scales such as in quark-gluon plasma \cite{Teaney:2001av}. It is therefore only natural to ask why such a simple set of equations works so well to explain reality across so many length scales. One popular explanation is that hydrodynamics is nothing but the gradient expansion of the conservation laws of a system, as elucidated in \cite{LLfluid}. 

In this note, we showcase linear instabilities that arise in the usual approach to hydrodynamics. These instabilities are represented as modes with a complex component that have a ``wrong'' sign, such that these modes grow exponentially in time. This is considered to be an unphysical instability, since physical fluids in homogeneous configuration are supposed to be linearly stable.

In order to fully appreciate the mentioned linear instabilities, we adopt a more formal perspective of fluids, with relativistic fluids as an example, in which one takes the viewpoint that hydrodynamics is an effective low-energy description of a theory with the following partition function 
\begin{equation}
  Z[g_{\mu \nu},A_\mu] = \left\la \exp\left[i \int d^{d+1}x\sqrt{-g}\left( T^{\mu \nu}g_{\mu \nu} + J^\mu A_\mu\right)  \right] \right\ra
  \,,
\end{equation}
where $g_{\mu \nu}$ is the background metric of the space where our ``fluid'' lives and the background gauge field $A_\mu$ represents a background (non-dynamical) electromagnetic field injected into the system. The independence of choice of coordinates and the gauge field implies that, in flat space with zero flux, there are the Ward identities 
\begin{equation}\label{eq:WardIdentity}
  \d_\mu T^{\mu}_{\;\;\,\nu} =0\,,\qquad \d_\mu J^\mu = 0\, .
\end{equation}
But just the conservation laws, by themselves, are not helping us to determine the evolution of the system. The next step is to assume that one can write down $T^{\mu}_{\;\;\,\nu}$ and $J^\mu$ in terms of macroscopic variables, such as local temperature $T(x)$, local chemical potential $\mu(x)$ and the fluid velocity $v^i(x)$. This process can be done, assuming that the system reaches thermal equilibrium in some infinitesimal volume, such that the local thermodynamic variables can be defined, even though the whole system is still evolving. The Noether currents $T^{\mu}_{\;\;\, \nu}$ and $J^\mu$ are then expressed in terms of these macroscopic proxies $\{T,\mu, v^i \}$, order by order in the derivative expansions, for example: 
\begin{equation}
\begin{aligned}
  T^{\mu}_{\;\;\,\nu} &= T^{\mu}_{0\;\nu} [\d^0] + T^{\mu}_{1\;\nu}[\d^1]+... \, ,\\
  \end{aligned}
\end{equation}
where the ellipsis denotes terms at higher-order in the derivative expansion. Due to the availability of extra tensor structures in the gradient expansion, transport coefficients are introduced. These are nothing but the coefficients in front of each independent structure. It is a common practice to simply include only first derivative terms to incorporate dissipative effects (such as shear and bulk viscosities) and ignore the higher-derivative terms. Once this is done, we have a system of equations that can be solved and is capable of determining the evolution of the system, namely,
\begin{equation}
  \d_\mu \left( T^{\mu}_{0\;\nu} + T^{\mu}_{1\;\nu} \right) = 0\, ,\qquad \d_\mu \left(J^\mu_0 + J^\mu_1  \right) = 0\, . \label{eq:1stOrderHydro-schematic}
\end{equation}
These equations are the relativistic analogue of the Navier-Stokes equations. On top of just writing down all possible structures $T^{\mu}_{1\;\nu}$ and $J^{\mu}_1$ constitute of, it is common to impose positivity of local entropy production. In practice, one constructs an entropy (density) current $s^\mu $, where $s^0$ in the static fluid  takes the form of local entropy density. Its divergence can be written schematically as 
\begin{equation}
  \partial_{\mu} s^\mu = \sum_i a_i X_{(i)\mu_1 \mu_2...}X^{\mu_1 \mu_2...}_{(i)} + \sum_j b_j X_{(i)\mu_1 \mu_2...}Y^{\mu_1 \mu_2...}_{(i)}+\CO(\d^3)\, , \label{eq:entropyProd}
\end{equation}
where $X^{\mu_1 \mu_2...}_{(i)}$ and $Y^{\mu_1 \mu_2...}_{(i)}$ are some first derivative tensors constructed out of $T$, $u^\mu$, $\mu$, with $X\neq Y$. Imposing $a_i \ge 0$ and $b_i =0$, which are functions of transport coefficients, guarantees that the entropy production is positive definite, i.e., $\partial_{\mu}s^{\mu}\geq0$. This places a restriction on the values of transport coefficients. Similar constraints can be found by imposing the local KMS conditions coming from the effective action approach in \cite{Crossley:2015evo}. See also \cite{Glorioso:2016gsa} for the proof of positivity of $\d_\mu s^\mu$ from unitarity of the microscopic constituents.
An additional principle that places restrictions on the values of transport coefficients are the Onsager relations, which are a direct result of assuming time reversal symmetry to hold in the theory \cite{Onsager:1931}.

There is, however, a big caveat in the above construction. Strictly speaking, $T$, $\mu$ and $v^{i}$, are uniquely defined only in equilibrium. It is nevertheless possible to redefine them by adding terms at higher-order in derivative expansion, e.g .,
%
\begin{equation}
  T(x) \to T(x) + \eth T+... 
  \,,
  \quad
  \mu(x) \to \mu(x) + \eth \mu+...
  \,,
  \quad
  v^{i}(x) \to v^{i}(x) + \eth v^{i}+...
  \,,
\end{equation}
where $\eth$ denotes objects that are exactly first order in derivatives and are constructed from $T, \mu ,v^i$.\footnote{The Icelandic letter $\eth$, pronounced $\textit{eth}$, can be used in latex via `\textbackslash eth'.} In principle, any choice, e.g., $T(x)$ or $T(x) + \eth T$, is an equally good macroscopic variable. The same applies to $\mu(x)$ and $v^i(x)$. The freedom of the formalism to choose macroscopic variables is usually referred to as \textit{frame choice} (see e.g. \cite{LLfluid,Kovtun:2012rj}). Chief among the popular frame choices are
\begin{itemize}
  \item \textit{Landau frame:} This is where the macroscopic variable is chosen such that 
\begin{equation}\label{eq:landau_frame}
T^{\mu}_{\;\;\,\nu} U^\nu = - \tilde{\CE}(T,\mu)U^\mu \, ,\qquad J^\mu U_\mu = n(T,\mu)
	\,,
\end{equation}
where $\tilde{\CE},n$ represent out-of-equilibrium energy density and $U(1)$ current density, which are functions of only $T(x)$, $\mu(x)$, $v^i(x)$ and \textit{not} their derivatives. We denote $U^\mu$ as a fluid 4-velocity constructed from $v^i$. Here, one can think of this frame as requiring the energy density flux to vanish.
  \item \textit{Eckart frame:} In this case, the variables are chosen such that 
\begin{equation}\label{eq:eckart_frame}
  T^{\mu}_{\;\;\,\nu}U_\mu U^\nu = -\tilde{\CE}(T,\mu) \, ,\qquad J^\mu = n(T,\mu) U^\mu
  \,,
\end{equation}
  where, again, $\tilde{\CE}$ and $n$ do not contain any derivatives of $T(x)$, $\mu(x)$, $v^i(x)$. Here, one can think of this frame as requiring the $U(1)$ density flux to vanish.
  \item \textit{General frame:} In this case, one picks $\delta T$, $\delta \mu$, such that $\tilde{\CE}$ and $n$ do not contain derivative corrections but the choice of $\eth v^i$ is not chosen to restrict $T^\mu_{\;\; \nu} U^\nu$ and $J^\mu$. This makes it possible to write down a superposition between Landau and Eckart frame.\footnote{Note that this notion of general frame is adopted from \cite{Hiscock:1985zz}. In some parts of the literature, the name general frame can refer to the case where none of the choices of $\eth T$ and $\eth \mu$ are made.}
\end{itemize}
While these choices  are used to simplify the equations of motion in Eq.~\eqref{eq:1stOrderHydro-schematic}, different choices result in different PDEs which, in principle, can yield very different solutions. 

In relativistic fluids, the consequence of the redefinition of these proxy macroscopic variables is very pronounced. This issue is demonstrated in the seminal work by Hiscock and Lindblom \cite{Hiscock:1985zz}, which studies linearised perturbations of both a stationary fluid and a fluid flowing at constant velocity, in all the frame choices mentioned above. Since these results are a crucial point in our note, let us summarise them here. Upon imposing $a_i \ge 0$ and $b_i =0$ in Eq.~\eqref{eq:entropyProd}, one finds that:
\begin{itemize}
  \item A linearised perturbation around first order relativistic hydrodynamics is unstable in the Eckart frame and in the general frame, even when the \textit{fluid is at rest}. On the other hand, the Landau frame is stable in this configuration.
  \item A linearised perturbation is unstable even in the Landau frame, when perturbed around \textit{a flow with constant velocity}.   
\end{itemize}
These linear instabilities invalidate first-order relativistic hydrodynamics as an effective theory. In the last few decades, numbers of proposals have been made to resolve this issue. Chief among them is to introduce new macroscopic degrees of freedom, which alter the structures of hydrodynamic modes at large frequency and wave vectors such as done in \cite{Muller:1967zza,Israel:1976tn,Israel:1979wp,Liu:1986,Geroch:1990bw}, for a review see e.g.~\cite{rezzolla2013relativistic}. Another populair approach is to view such unstable modes as an artifact of the truncation of the gradient expansion \cite{Baier:2007ix,Grozdanov:2015kqa,Heller:2013fn,Withers:2018srf,Grozdanov:2019kge}. The former is motivated by kinetic theory of gases \cite{grad-1963}, while the latter emerges from the realisation of the hydrodynamical limit in a strongly coupled quantum field theory with a holographic dual \cite{Policastro:2002se,Bhattacharyya:2008jc}. We will return to some of these proposals in the discussion of section \ref{sec:summary}, as well as other open issues. Nevertheless, while these resolutions are available and possibly correct (albeit in different regimes), one thing is clear: first order relativistic hydrodynamics, as presented, is not a good description for fluids due to its linear instabilities.

 One obvious question to ask is  whether these artificial instabilities are specific to a fluid with Lorentz boost symmetry or whether they are symptoms of the truncation of the gradient expansion. In this note, we would like to present evidences toward the latter case. 
 Particularly, we show that first order fluids are unstable even when we replace the algebras of $T^{\mu}_{\;\;\,\nu}$ and $J^\mu$ from Poincar\'e + $U(1)$ with the Bargmann algebra or with the Carrollian + $U(1)$ algebra. The former is the usual massive Galilean non-relativistic fluid (as we will see, this algebra is obtained in a more subtle way than just a naive $c \to \infty$ contraction of Poincar\'e + $U(1)$ algebra!), while the latter corresponds to the contraction $c\to 0$, which is sometimes referred to as ultra-relativistic limit. Because of these potentially confusing naming conventions, we shall refer to these two types of fluids as non-Lorentzian fluids. We found that the first order Carrollian fluid shares the same symptoms as the Lorentzian fluid, as found in \cite{Hiscock:1985zz}. As for the Bargmann fluid, contradicting to what is commonly believed, it turns out that not all ``non-relativistic fluids'' are stable. Namely, there are frame choices where the Bargmann fluid also suffers from the spurious instability in the same way as the Lorentzian fluid. It turns out that only the Eckart frame of a Bargmann fluid is stable.

Additionally, in section \ref{generalfluid}, we show that for a generic fluid without imposing any boost symmetry, it is also possible that such a fluid exhibits the same kind of instabilities as \cite{Hiscock:1985zz}. We furthermore present conditions on transport coefficients for which the static first order fluid without boost symmetry is stable. This analysis can be thought of as a guideline for making a frame choice and extensions of hydrodynamics (which we briefly discussed above and will further elaborate in Section \ref{sec:summary}) to remove such spurious unstable modes.

 Other than investigating the nature of the unstable modes of the Lorentzian and Bargmann fluids, it is beneficial to study the spectrum of the Carrollian fluid \cite{Ciambelli:2018wre,Ciambelli:2018xat,deBoer:2017ing}. Studying such a fluid can be relevant in the context of flat space holography, due to the connection between the Carrollian group and the BMS group \cite{Duval:2014uva,Duval:2014uoa}. See also \cite{Bagchi:2016bcd} and references therein. Furthermore, it was argued that the membrane paradigm can be cast into a Carrollian fluid formulation \cite{Donnay:2019jiz}. To the best of our knowledge, little is known about the spectrum of hydrodynamic modes in Carrollian fluids.  

 We summarise our findings in Section \ref{sec:summary}: which frame choice of what algebra is unstable under what circumstances. We also discuss in what way these unstable modes could be removed. Section \ref{sec:technicalOverview} provides an overview of the technical details concerning thermodynamical variables and frame choices, while Section \ref{sec:bargmann} and \ref{sec:carroll} are dedicated to computing of the spectra of Bargmann and Carrollian fluids in different frame choices. Finally, in Section \ref{generalfluid}, we analyse the spectrum of a static fluid without imposing any boost symmetry, in various frames.
\section{Summary and outlook}\label{sec:summary}

We emphasise that we are not claiming that hydrodynamics (which supposedly describes real fluids) is unstable. Our goal is to point out that (i) expressing $T^{\mu}_{\;\;\,\nu}$ and $J^\mu$ in terms of $\{ T, \mu, v^\mu\}$ and their first derivatives and (ii) imposing the \textit{strict} positivity of $\nabla_\mu s^\mu$, namely $a_i \ge 0$ and $b_i =0$, implies that \textit{first order} hydrodynamics contains linear instabilities beyond Lorentzian symmetry. Many of the controversial sounding statements have already been addressed, for decades, in the literature. The results concerning instabilities of Lorentzian fluids, constructed via principles (i) and (ii), are well-known \cite{Hiscock:1985zz,rezzolla2013relativistic}. The theory of non-relativistic fluids, which has an even longer history, presented as a gradient expansion of conserved currents can be found in, e.g., \cite{LLfluid,Son:2005rv} and particularly \cite{Jensen:2014ama}. The construction of a Carrollian fluid, on the other hand, has been considered very recently \cite{Ciambelli:2018wre,Ciambelli:2018xat,deBoer:2017ing}. Also, only not long ago a beginning was made in studying fluids with relaxed boost symmetry conditions \cite{deBoer:2017ing,deBoer:2017abi}. Of course, there are many more developments of these fluids that we can hardly do justice. What we did is simply analysing linear perturbations of these non-Lorentzian fluids and show that they contain the same kind of artificial unstable modes as in the Lorentzian case.

\begin{figure}[b]
\centering
\includegraphics[width=0.35\textwidth]{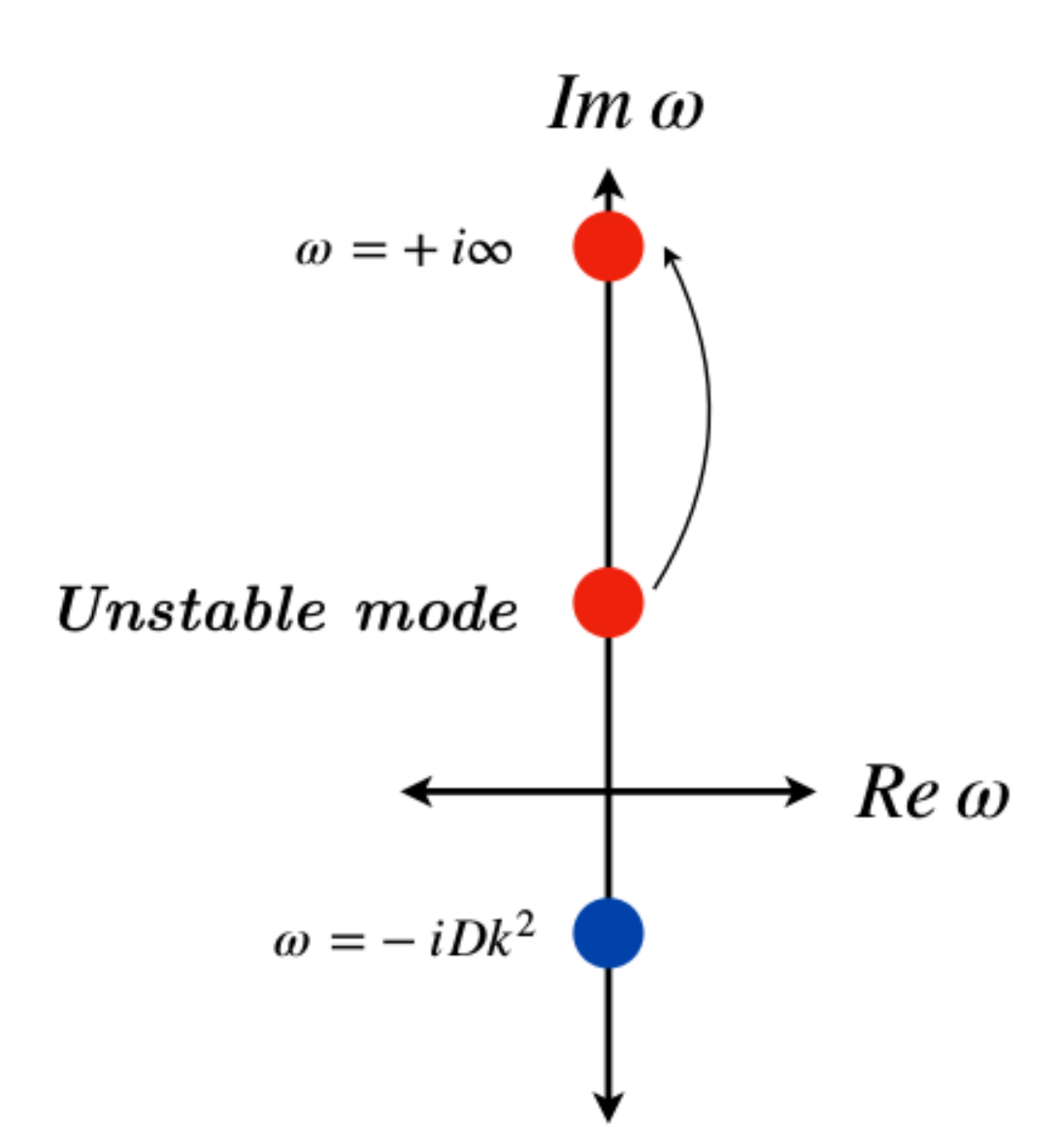}
\includegraphics[width=0.35\textwidth]{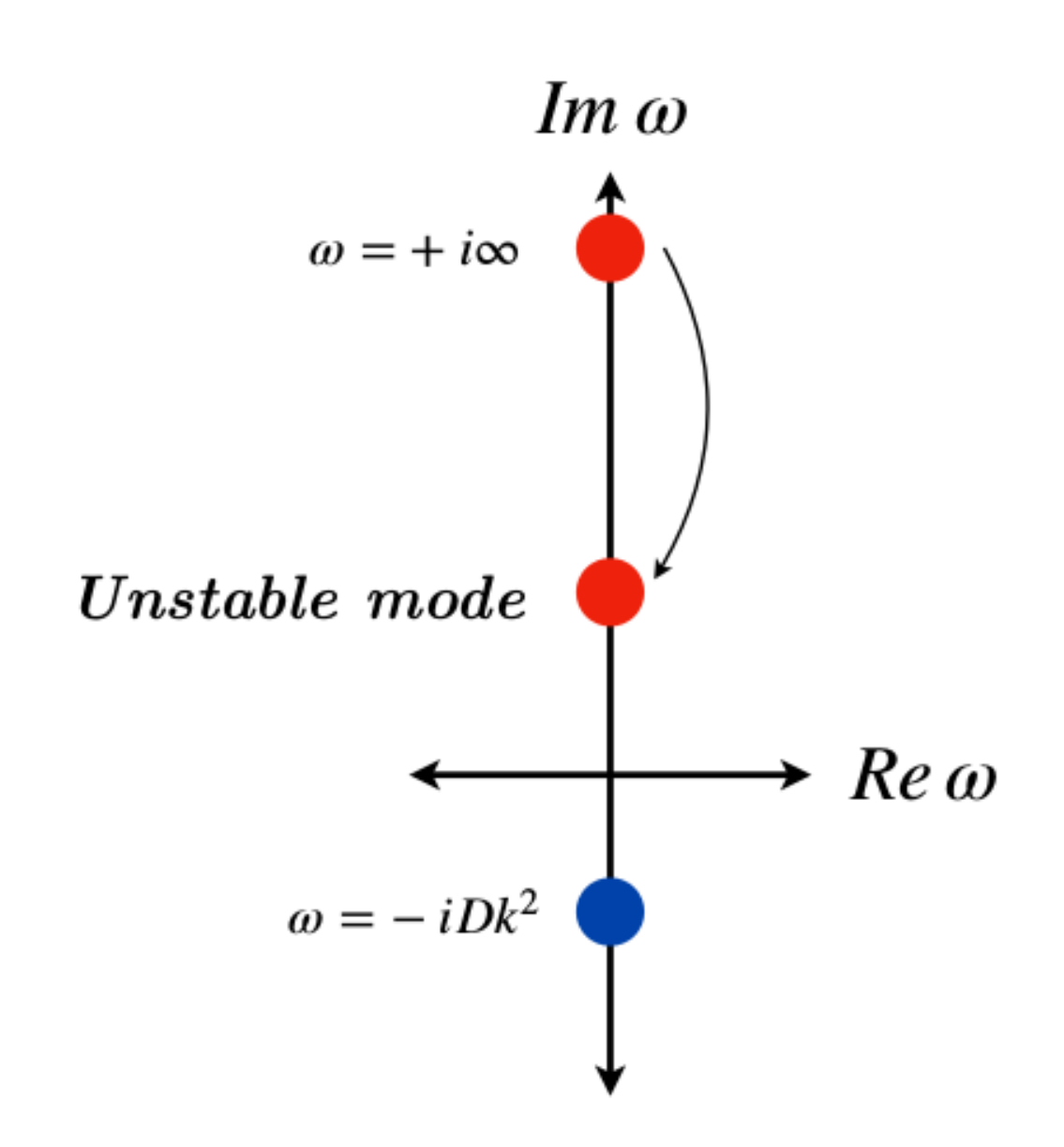}
\captionsetup{justification=raggedright,
singlelinecheck=false
}
\caption{\textbf{(LEFT)} The pole structure of the transverse fluctuations around the static $(P_i =0)$, homogeneous configuration of a Lorentzian, Bargmann or Carollian fluid in general frame. The unstable mode is moved towards $\omega \to +i\infty$ and is removed from the spectrum as one continuously tunes the transport coefficient towards the Landau frame (for Lorentzian and Carrollian fluid) and Eckart frame (for Bargmann fluid). \textbf{(RIGHT)} This panel illustrates the unstable mode in the Landau frame of a Carrollian or a Lorentzian fluid at $P_i \ne 0$, which moves down from $\omega \to +i\infty$ as we move away from $P_i = 0$ configuration. This pole is absent in the Eckart frame of the Bargmann fluid. }
\label{fig:Poles}
\end{figure}

Let us now summarise what happens if we make different frame choices. In some frames, the theory is outright unstable, see Figure \ref{fig:Poles} (LEFT). It contains modes in the lower-half complex $\omega$ plane as well as poles in the upper-half plane. The latter poles exist even when the wave vector $k^i = 0$. If such poles are located in the lower-half plane, they will decay away in the hydrodynamic regime (late time and long distance). However, the pole in the upper-half plane is an indication of an instability of the theory. In some frame, this unstable mode disappears when the fluid is at rest or, more specifically, when one studies the perturbation around the fluid with vanishing momentum density of the fluid, $P_i:= T^0_{\;\;i}=0$. However, it becomes unstable when $P_{i}$ is nonzero, see Figure \ref{fig:Poles} (RIGHT). Table \ref{fig:table} below, summarises the scenario in which unstable modes occur in a fluid with Lorentzian + $U(1)$, Bargmann, or Carrollian + $U(1)$ as its symmetries in both Eckart and Landau frame.
\begin{figure}
\begin{center}
\begin{tabular}{ |p{3cm}||p{2.4cm}|p{2.4cm}| p{2.2cm}| p{2.2cm}|  }
 \hline
 & \text{Landau frame} $P_i =0$& \text{Landau frame} $P_i \ne 0$ &\text{Eckart frame} $P_i =0$ & \text{Eckart frame} $P_i \ne0$\\
 \hline
 Lorentzian + $U(1)$& Stable & \textit{Unstable} & \textit{Unstable} & \textit{Unstable} \\ 
 \hline
 Bargmann & \textit{Unstable} & \textit{Unstable} & Stable & Stable \\
 \hline
 Lorentzian + $U(1)$
  $c\to 0$ limit & Stable & \textit{Unstable} & \textit{Unstable} & \textit{Unstable} \\
 \hline
 \end{tabular}
 \vspace{-0.4cm}
\end{center}
\captionsetup{justification=raggedright,
singlelinecheck=false
}
\caption{In this table, we showcase an overview of instabilities of first order hydrodynamics, including our results. Here $P_{i}:=T^{0}_{\;\;\,i}$ denotes momentum density, which can be made zero or non-zero by performing a boost. \vspace{-0.5cm}}\label{fig:table}
\end{figure}

We furthermore show, that if one relaxes the constraint of boost symmetry, there are still possibilities for instabilities in a static fluid. One can use our results as a guideline for putting restrictions on values of transport coefficients for which, at least for a static fluid, the instability is absent.

As for the Carrollian fluid, we can trace the instability from the Lorentzian fluid via the $c\to 0$ contraction (see Section \ref{sec:carroll}). Another subtle issue of the Carrollian fluid that we found, is the lack of sound modes up to first order, even though the theory is invariant under translations. This can be seen from the fact that the susceptibility of the energy density and the momentum density (as well as of energy density and $U(1)$ density) vanishes. The collective excitations, therefore, turn out be diffusive modes. Furthermore, we found that simply taking the $c\to0$ contraction of both frames of a Lorentzian fluid, does give rise to Carrollian fluids, but they seem to be non-connected by a frame choice. In other words, they might both be different manifestations of a Carrollian fluid. This seems to reflect the fact that  the longitudinal fluctuation, in the one obtained via the $c\to 0$ limit of the Lorentzian fluid in the Landau frame, exhibits two diffusive modes, while the one obtained via the $c\to 0$ limit of the Lorentzian fluid in the Eckart frame only contains one diffusive mode. Since diffusive modes are physical quantities, it is an indication that choosing different frame choices \textit{before} taking the $c\to 0$ limit results in two different theories. It would be interesting to better understand this fact from the construction of Carroll fluids without using $c\to0$ contractions.

One thing we should note is the fact that the choice of out-of equilibrium proxies $T,\mu,v^i$ is called \textit{frame choice}, is rather misleading. The word \textit{frame choice} sounds as if it has something to do with the reference frame of a certain boost symmetry. We have already seen from the current note, that it has nothing to do with boost symmetry. Moreover, it sounds as if it is our choice to choose whichever frame we want to simplify a computation. For first order hydrodynamics, different frame choices can lead to (different) instabilities. One could argue that a reasonable effective description of a fluid, should be formulated in a frame choice in which it makes the system, at least, linearly stable. However, away from the practical point of view, there is no fundamental principle that can distinguish one frame choice from another. It would be very interesting to find a microscopic way to derive these out-of-equilibrium macroscopic proxies in a way that the frame choice would be completely determined. In other words, when we obtain hydrodynamics from truncating higher spin currents (as in kinetic theory) or from truncating the infinite series (as in holography), is it possible to consistently determine \textit{a good definition of macroscopic variables}? To the best of our knowledge, this issue remains unresolved.

What do all of these analyses mean in a bigger picture? One question at the back of our minds is \textit{why does hydrodynamics work so well}? A conventional answer is: because hydrodynamics is a gradient expansions of Noether currents in terms of macroscopic quantities. But as we have already seen here, as well as in previous works, choosing some version of macroscopic variables leads to artificial instabilities. While these phenomena are well-documented in the Lorentzian fluid, we show that such instabilities can occur beyond the Lorentzian fluid in first order hydrodynamics. The exception that we found is in the non-relativistic (Bargmann) fluid in the Eckart frame. The absence of the spurious instability in the Eckart frame of the Bargmann fluid makes sense, since it is nothing else than the famous and well tested Navier-Stokes equations. But what about other incarnations of first order hydrodynamics? Can they still serve as good effective descriptions, despite their instabilities? While we, regrettably, cannot provide a conclusive answer, there are several proposals concerning the modification of first order hydrodynamics in the Lorentzian case, where the artificial instability is well investigated. Let us review some these possibilities below: 

\begin{itemize}
  \item \textit{M\"uller-Israel-Stewart (MIS) type theory:} Inspired by kinetic theory,\footnote{See e.g. \cite{Romatschke:2009im} for a derivation of a theory similar to MIS theory from (truncated) higher-spin currents in kinetic theory. A brief recent discussion on the role of higher-spin currents in the kinetic theory of gases can also be found in \cite{Grozdanov:2018fic}.} one way to formulate a theory that is free of the mentioned unphysical instabilities is to reformulate the theory in the Landau frame and promote the dissipative part $t^{\mu}_{\;\;\,\nu}$ of the stress-energy tensor to a full-fledged dynamic field (see e.g. Eq.~\eqref{eq:general-Lorentzian-constitutiveReln} of Section \ref{sec:technicalOverview} for more details on the conventions). It is then required to provide additional equations of motion beyond the conservation law for $t^{\mu}_{\;\;\,\nu}$, see e.g. \cite{Muller:1967zza,Israel:1976tn,Israel:1979wp}. While being useful phenomenologically, its physical origin is rather unclear.  Moreover, such equations are not unique as seen in \cite{Liu:1986,Geroch:1990bw}, where the additional equation is modified in order to ensure causality and hyperbolicity (see also \cite{rezzolla2013relativistic} for review). It was recently argued in \cite{Grozdanov:2018fic}, from the point of view of memory matrix formalism, that theories described by MIS theory contain an additional slightly broken symmetry, where the equations of motion of $t^{\mu}_{\;\;\,\nu}$ can be thought of as the almost conservation law. If true, it could justify the presence of the added degrees of freedom and their equations of motion. While these possibilities have been explored for the Lorentzian fluid, we are not aware of works in this direction for non-Lorentzian first order hydrodynamics.  

  \item \textit{Hydrodynamics as a gradient expansion:} In strongly interacting systems,
the gauge/gravity duality provides a description of hydrodynamics as an infinite series of gradient expansions. From the dual gravity side, the artificial instability is absent and thus implies that hydrodynamics as a gradient expansion should also be stable. This strongly suggests that the instability of Hiscock-Lindblom type originates from truncating the infinite gradient expansion series. Of course, one will immediately ask whether there is a way to consistently truncate the series and, if the truncation to first derivative is problematic, how many orders in the derivative expansion we should keep. It would be interesting to use the gauge/gravity duality, particularly the method to extract nonlinear constitutive relations from a gravity dual such as \cite{Bhattacharyya:2008jc,Erdmenger:2008rm,Banerjee:2008th,Rangamani:2008gi,Brattan:2010bw,Kiritsis:2015doa,Kiritsis:2016rcb}, to gain more insight into these questions. 

	\item \textit{The entropy current:} Recently it was shown that there is a way of making first order uncharged Lorentzian hydrodynamics stable, evading the shortcoming of the constructions mentioned above \cite{Van:2011yn,Kovtun:2019hdm,Bemfica:2019knx}. The authors of \cite{Van:2011yn,Kovtun:2019hdm,Bemfica:2019knx} argued that there is a class of frame choices (neither Eckart or Landau frame), where $\tilde \CE$ and $T^\mu_{\;\;\,\nu}U^\nu$ receive derivative corrections, which are linearly stable. Furthermore, as particularly emphasised in \cite{Kovtun:2019hdm}, the constraint on the entropy current is relaxed to only the configuration that obeys the equations of motion. This approach constrains the transport coefficients considerably less than the approach used in \cite{Hiscock:1985zz} (we shall review this construction shortly in Section \ref{sec:technicalOverview}). It would be interesting to understand if there is a physical reasons, other than the fact that they are stable, why these frame choices are the preferred one. 

\end{itemize}

Furthermore, we would like to point out the role of the Carrollian fluid as a potential dual description of gravity theory in asymptotically flat spacetime \cite{Duval:2014uva,Duval:2014uoa,Bagchi:2016bcd}. When one would be able to construct an equivalent of the well studied AdS/CFT fluid-gravity correspondence, see \cite{Rangamani:2009xk} for a review, one could imaging using Carrollian fluids as an input to learn about the dynamics of flat spacetime (or vice versa). We contribute to this avenue of research by showing that a Carroll fluid in Landau frame is, from the instability point of view, as trust-worthy as the relativistic fluid in Landau frame (which is employed in the context of fluid-gravity). The absence of the sound mode, at least up to first order, could be an interesting fact in the light of this construction.

Finally, in the context of the membrane paradigm, it is suggested that some stretched membrane near the horizon of a black hole experiences the emergence of Carrollian symmetries \cite{Donnay:2019jiz}. In particular this would imply that a fluid description of the dynamics on such a membrane, at least at leading order, is governed by a Carrollian fluid. More specifically, the here established results should be found to be encoded in the horizon dynamics of a black hole.
\paragraph*{NOTE ADDED:}
The work of \cite{novak2019hydrodynamics}, appeared in parallel, also discussed frame choices of the fluid without boost which overlap with Section \ref{sec:general} and Appendix \ref{appendix1} of our manuscript. Furthermore, Ref.\cite{novak2019hydrodynamics} also extend the constitutive relations for such fluid, from the case where fluid is at rest (discussed in Section \ref{sec:general}), to the case where it attains finite velocity.

\section{Technical overview}\label{sec:technicalOverview}

The goal of this section is to review important concepts and aspects of different fluids, while establishing notation. We will start by considering the relativiitfluid with relaxed boost constraints. 
\subsection{Relativistic fluid}
Let us consider a relativistic fluid with some $U(1)$ charge. Here, relativistic refers to the fact that the fluid has symmetries according to the Poincar\'{e} group. We introduce fluid velocity $v^{i}$, which we will incorporate into (covariant) fluid velocity $U^{\mu}$. In order to expand our constitutive relations perpendicular and orthogonal to $U^{\mu}$, we construct projector $\Delta^{\mu}_{\;\;\,\nu}$. We explicitly take
\begin{equation}
	U^{\mu}
	:=
	\frac{(1,v^{i})}{\sqrt{1-\frac{v^{2}}{c^{2}}}}
	\,,
	\quad
	\Delta^{\mu}_{\;\;\,\nu}
	:=
	\delta^{\mu}_{\;\,\nu}
	+
	\frac{U^{\mu}U_{\nu}}{c^{2}}
	\,.
\end{equation}
In order to raise and lower indices we use the metric $g_{\mu\nu}$ for the Lorentzian fluid. Note that we will work in the convention where $U^\mu U_\mu = -c^2$ and that the metric in flat space is 
\begin{equation}
  g_{\mu \nu} = -c^2 dt^2 + \delta_{ij}dx^i dx^j \,, \qquad 
\end{equation}
where $i,j = 1,2,3,..,d$ labels the spatial direction.
Projecting perpendicular and parallel to $U^{\mu}$, we find that the constitutive relations of a relativistic fluid, with some $U(1)$ charge, can generically be written as
\begin{equation}\label{eq:general-Lorentzian-constitutiveReln}
\begin{aligned}
	T^{\mu}_{\;\;\,\nu}
	=&\;
	\tilde{E}U^{\mu}U_{\nu}
	+
	\mathcal{P}\Delta^{\mu}_{\;\;\,\nu}
	+
	(Q^{\mu}U_{\nu}+U^{\mu}Q_{\nu})
	+
	t^{\mu}_{\;\;\,\nu}
	\,,
	\\
	J^{\mu}
	=&\;
	\mathcal{N}U^{\mu}
	+
	j^{\mu}
	\,.
\end{aligned}
\end{equation}
Here $\tilde{E}$, $\mathcal{P}$, $\mathcal{N}$ are scalars, the vectors $j^{\mu}$ and $q^{\mu}$ are transverse (to $U_{\mu}$) and the tensor $t^{\mu}_{\;\;\nu}$ is transverse, symmetric and traceless. These scalars, vectors and tensor are, in principle, functions of all possible (derivatives of) $U^{\mu}$, temperature $T$ and chemical potential $\mu$. 

Temperature and chemical potential are related to pressure $P$ and internal energy density $\tilde{\mathcal{E}}$ via the following first law and Euler relation\footnote{When considering thermodynamics we adopt the notational conventions of \cite{deBoer:2017ing}.}
\begin{equation}
	d\tilde{\mathcal{E}}
	=
	Tds
	+
	\mu d\tilde{n}
	\,,
	\quad
	\tilde{\mathcal{E}}
	+
	P
	=
	T
	s
	+
	\mu
	\tilde{n}
	\,.
\end{equation}
At the level of the perfect fluid, the constitutive relations receive (by construction) no derivative corrections and thus $\tilde{E}=\tilde{\mathcal{E}}/c^{2}$, $\mathcal{P}=P$, $N=\tilde{n}$ and $q^{\mu}=j^{\mu}=t^{\mu}_{\;\;\,\nu}=0$.
Once one has obtained a closed form of the constitutive relations like in the perfect fluid case above, one can compute the equations of motion for a fluid, which are defined by $\partial_{\mu}T^{\mu}_{\;\;\,\nu}=0$, $\partial_{\mu}J^{\mu}=0$.

We are interested in fluids with first order corrections in derivatives. As discussed in the introduction, see the text surrounding equations Eq.~\eqref{eq:landau_frame} and Eq.~\eqref{eq:eckart_frame}, one has to adopt a frame choice and require for the entropy current $s^{\mu}$ for which $\partial_{\mu}s^{\mu}\geq0$ for all fluid configurations. All these results combined, results in following expressions for $\tilde E,\CP, \CN$ and $t^\mu_{\;\; \nu}$ for all considered frames \cite{Kovtun:2012rj}
\begin{equation}\label{eq:Lorentz-derivatives corrections}
\begin{aligned}
	\tilde{E}&=\frac{\tilde{\mathcal{E}}}{c^{2}}
	\,,
	\quad
	\mathcal{P}
	=
	P
	-
	\zeta\partial_{\lambda}U^{\lambda}
	\,,\quad
	\mathcal{N}=\tilde{n}
	\,,\\
	t^{\mu}_{\;\;\,\nu}
	&=
	-\eta \Sigma^{\mu}_{\;\;\,\nu}
	:=-
	\eta\Delta^{\mu\alpha}\Delta^{\beta}_{\;\;\,\nu}
	\left( \partial_{\alpha}U_{\beta}
		+
		\partial_{\beta}U_{\alpha}
		-
		\frac{2}{d}\eta_{\alpha\beta}\partial_{\lambda}U^{\lambda}\right)\,.
  \end{aligned}
\end{equation}
Depending on frame, the vectors $q^{\mu}$ and $j^{\mu}$ are different. For the \textbf{Eckart frame}, see Eq.~\eqref{eq:eckart_frame}, one finds the requirement
\begin{equation}
	Q^{\mu}=-\kappa\Delta^{\mu\nu}(TU^{\lambda}\partial_{\lambda}U_{\nu}+\partial_{\nu}T)
	\,,
	\quad
	j^{\mu}
	=
	0
	\,,
\end{equation}
whereas the \textbf{Landau frame}, see Eq.~\eqref{eq:landau_frame}, requires
\begin{equation}
	Q^{\mu}=0
	\,,
	\quad
	j^{\mu}
	=
	-
	\sigma T\Delta^{\mu\nu}\partial_{\nu}\frac{\mu}{T}
	\,.
\end{equation}
Here the transport coefficients $\eta,\zeta,\sigma$ are positive definite. The heat conductivity can be related to charge conductivity via $\kappa=\sigma(\tilde{\mathcal{E}}+P)/(\tilde{n}^{2}T)$. The \textbf{general frame} is a superposition of the two frames above, where there is no more relation between $\sigma$ and $\kappa$. In \cite{Hiscock:1985zz}, it was argued that the divergence of the entropy production can be expressed as 
\begin{equation}
 T \d_\mu s^\mu = - t^{\mu \nu} \sigma_{\mu \nu}   -(\CP-P) \d_\lambda U^\lambda-Q^\mu \Delta_{\mu \nu} \left( T  U^\lambda \d_\lambda U^\nu + \d^\nu T  \right) - j^\mu \d_\mu(\mu /T)
 \,,
\end{equation}
where $T s^\mu = P U^\mu - T^{\mu \nu}U_\nu - \mu J^\mu = s U^\mu +\text{(first derivative correction)}$. It is then concluded that for $\CP$ and $t^{\mu \nu}$ in \eqref{eq:Lorentz-derivatives corrections} with 
\begin{equation}
  Q^\mu = -\kappa T \Delta^{\mu \nu} \left( T U^\lambda\d_\lambda U_\nu + \frac{\d_\nu T}{T} \right)\, ,\qquad j^\mu = - \sigma T \nabla^{\mu \nu} \d_\nu \left( \frac{\mu}{T} \right)\, ,
\end{equation}
where the transport coefficients are constrained to be 
\begin{equation}
  \eta \ge 0 \, , \qquad \zeta \ge 0 \, , \qquad \sigma \ge 0 \, ,\qquad \kappa \ge 0\,.
\end{equation}
The above conditions are guaranteed to give positive entropy production in any fluid configuration.\footnote{It is argued, see e.g. Appendix A of Ref.\cite{Kovtun:2019hdm}, that this constraint is too strong and that, upon restricting to on-shell configurations, the positivity of $\d_\mu s^\mu$ provides less constraining conditions on the transport coefficients.} These constraints will be inherited by non-Lorentzian fluids that can be obtained through the $c\to \infty$ and the $c\to 0$ limit of the Lorentzian fluid. 

The linear stability analysis of Eckart, Landau and general frame of this theory has been performed in \cite{Hiscock:1985zz}. We already summarised their results in previous sections and will therefore not repeat it here. 

\subsection{From Lorentzian to Bargmann and Carrollian fluids}\label{sec:technicalOverview-Bargmann}

In this section, we will give a summary of the procedure on how to obtain the constitutive relations of Bargmann and Carrollian fluids, from the Lorentzian one. The steps presented here are not new, but it can be useful for comprehending different frames in different fluids.

Firstly, it proves useful to write down the metric in such a way that the time direction is apparent. Namely,
\begin{equation}\label{eq:nonrelativisticMetric}
  g_{\mu \nu} = - c^2 \bar\tau_\mu \bar\tau_\nu + h_{\mu \nu} \, , \qquad g^{\mu \nu} = -\frac{1}{c^2}\tau^\mu \tau^\nu+ h^{\mu \nu}\,,
\end{equation}
where $\tau^\mu$, $\bar\tau_\mu$, $h^{\mu \nu}$, $h_{\mu \nu}$ satisfy the orthogonality conditions
\begin{equation}
  \tau^\mu h_{\mu \nu} = 0\, , \qquad \bar\tau_\mu h^{\mu \nu} = 0\, ,\quad \tau^\mu \bar\tau_\mu = 1\,,
\end{equation}
and the projector to the plane orthogonal to $\{\tau^\mu,\bar\tau_\mu\}$ is $P^\mu_{\;\;\, \nu} := h^{\mu \rho} h_{\rho \nu}$. Why are we doing this, instead of simply taking the flat Minkowski space limit and performing $c\to \infty$ (to Bargmann) or $c\to 0$ (to Carrollian)? Firstly, it is useful to keep the generic metric, in order to define a generating function. Secondly, this notation allows us to take the limit in a covariant way, while being able to keep track of the transformations of the geometric quantities, such that is ensured that the Bargmann and Carrollian algebra are manifest in these fluids.

Let us start with the $c\to \infty$ limit, in order to obtain the Bargmann fluid constitutive relations. Afterwards, we proceed to obtain the Carrollian fluid constitutive relations. While the former metric was introduced in general form, our computations will be done only in flat space, where 
\begin{equation}
  \tau^\mu = \delta^\mu_{\;\;0}\, , \qquad \bar\tau_\mu = \delta^0_{\;\;\mu}\, ,\qquad u^\mu = (1,v^i)\, ,\qquad h_{\mu \nu} = \text{diag}(0,1,1,1)\, ,
\end{equation}
which might be useful in order to visualise the computation.

\subsubsection{Bargmann fluid}
Let us first start with the fluid velocity. It is convenient to decompose the Lorentz covariant velocity as 
\begin{equation}
  U^\mu = \gamma u^\mu \, ,\qquad u_\mu = h_{\mu \nu}u^\nu \, ,\qquad \text{where} \qquad \lim_{c\to \infty} U^\mu = u^\mu\,,
\end{equation}
and where the velocity $u^\mu$ is no longer normalised to speed of light, i.e., $u^2 := u_\mu u^\mu \ne -c^2$ but instead $u^2 = v^2$. Note that $\bar\tau_\mu u^\mu = 1$, but $\tau^\mu u_\mu = \tau^\mu h_{\mu \nu} u^\nu = 0$. After some algebra, we obtain the following $c\to \infty$ limit of the (Lorentzian) projectors 
\begin{equation}
\begin{aligned}
  \lim_{c\to \infty}\Delta^{\mu \nu} &= h^{\mu \nu} \, ,\\
   \lim_{c\to \infty}\Delta^{\mu}_{\;\;\,\nu} &= h^{\mu \rho} \left(h_{\nu \rho} - \bar\tau_{\nu}u_{\rho}- \bar\tau_{\rho}u_{\nu}  + \bar\tau_\nu \bar\tau_\rho u^2  \right) \, ,\\
   \lim_{c\to \infty}\Delta_{\mu \nu} &= h_{\mu \nu} - \bar\tau_{\mu}u_{\nu}- \bar\tau_{\nu}u_{\mu} + \bar\tau_\mu \bar\tau_\nu u^2  \, .
\end{aligned}
\end{equation}
Note the influence of factors of $c$ in the metric tensor.

We might think that by simply taking $c\to \infty$ of the Lorentz constituitive relations, we will obtain the Bargmann constituitive relations on the nose, but there are some subtle issues there. To start, a naive $c\to \infty$ limit of the Lorentzian fluid will give us a vanishing momentum density flux: $P_i := T^0_{\;\;\,i} = 0$, see e.g. \cite{deBoer:2017ing}. This is in conjunction with the fact that the boost generator $Q_i$ and generator of translations $T_i$ commute in the naive $c\to \infty$ limit of the Poincar\'{e} algebra, see e.g. \cite{Andringa:2010it}, implying that a fluid at rest cannot be boosted to finite momentum. This is clearly an unnatural feature for a non-relativistic fluid in everyday life and in this limit, one ought to modify the commutator 
\begin{equation}\nonumber
[Q^i,T_j] = 0\,,\qquad \qquad \text{to} \qquad \qquad [Q^i,T_j] = -\delta^i_{\;\,j} m N\,,
\end{equation}
which results in what is known as the Bargmann algebra. The parameter $m$ plays the role of mass in the non-relativistic theory and $N$ is the generator of the $U(1)$ central extension. It will turn out that due to this central extension, boosting the fluid will cause the momentum to change accordingly. In the context of the $c\to\infty$ contraction, the modification of the commutator above arises from the choice of considering a term on the right hand side that is otherwise suppressed.

The above limit where $P_i = 0$ (or equivalently, $m=0$), is referred to as \textit{Massless Galilean} fluid \cite{deBoer:2017ing} and to get away from it at the level of constitutive relations, we first note the ambiguity of the $c\to\infty$ limit of the metric written in the form of Eq. \eqref{eq:nonrelativisticMetric}. That is, we can perform the following transformation as $c\to \infty$ 
\begin{equation}
  \bar\tau_\mu \to \bar\tau_\mu - \frac{1}{c^2}\Psi_\mu \,, \qquad h_{\mu \nu}\to h_{\mu \nu} - 2 \bar\tau_{(\mu} \Psi_{\nu)} + \frac{1}{c^2} \Psi_\mu \Psi_\nu  
  \,,
\end{equation}
which also shifts the corresponding background $U(1)$ gauge field, attributed to the central extension, by $A_\mu \to A_\mu + m\Psi_\mu$. Here $\Psi_\mu$ is an $\CO(c^0)$ quantity and the $c\to\infty$ limit, implies that only $A_\mu$ and $h_{\mu \nu}$ transform.\footnote{Note also that in order to ensure that $h_{\mu \nu}$ remains a rank $d$ tensor ($d=3$ in our case), we need $\Psi_\mu = \psi_\mu - \frac{1}{2} \bar\tau_\mu \psi^2$ where $\psi_\mu \tau^\mu = 0$ and $\psi^2 = h^{\mu \nu} \psi_\mu \psi_\nu$. See also Section 2 of \cite{Jensen:2014wha} for a thorough discussion.} This shift symmetry is called Milne boost \cite{Duval:1983pb} and ensures a relation between the $U(1)$ current $J^i$ and the momentum density flux $P_i$, see also \cite{Jensen:2014aia,Jensen:2014ama,Jensen:2014wha} for a recent discussion, where most material in this subsection is based upon. The Milne boost can be thought of as a geometric implementation of the Galilean boost.

Now, we finally have the machinery to take the $c\to \infty$ limit properly. It turns out that by directly taking the $c\to\infty$ limit of the stress-energy tensor, we find the Milne boost invariant object $\CT^{\mu \nu}$, defined via 
\begin{equation}\label{eq:Milne-invariant-T}
  \CT^{\mu \nu} = \lim_{c\to \infty} c\, T^{\mu \nu}_\text{Lorentz}  \,.
\end{equation}
Milne boost invariance implies furthermore
\begin{equation}\label{eq:Milne-wardIdentity}
  J^\mu := m \lim_{c\to \infty} c\,J^\mu_\text{Lorentz} =  \CT^{\mu \nu} \bar \tau_\nu \, ,\qquad P_\mu = J^\nu h_{\mu \nu}\, .
\end{equation}
For all practical purposes, we can think of the above relation as a definition of the $U(1)$ current and the momentum density current. The other Milne boost invariant object that is crucial to this theory, is the Milne boost invariant energy density
\begin{equation}\label{eq:Milne-invariant-E}
  \tilde \CE^\mu = -\lim_{c\to \infty} \left( T^{\mu \nu}_\text{Lorentz} U_\nu + mc^2\, J^\mu_\text{Lorentz} \right)
  \,.
\end{equation}
Note that this is not the total energy density of the system as the latter is not invariant under a Milne boost. The Milne Ward identity \eqref{eq:Milne-wardIdentity} combined with $\tilde \CE^\mu$ and $\CT^{\mu \nu}$, serves as macroscopic data, analogous to $T^\mu_{\;\;\,\nu}$ and $J^\mu$ in the Lorentzian case. 

From the above, it follows that regularity in the $c\to\infty$ limit alters our definition of frame choice between Lorentz and Bargmann. Substituting the constitutive relation in Eq. \eqref{eq:general-Lorentzian-constitutiveReln} into the Milne invariant current \eqref{eq:Milne-invariant-T} and \eqref{eq:Milne-invariant-E}, we find that 
\begin{equation}\label{eq:Milne-invariant-constitutiveReln}
  \begin{aligned}
\tilde \CE^\mu &= \tilde \CE u^\mu + \eta^\mu \, ,\\
\CT^{\mu \nu} &=  n u^\mu u^\nu + P h^{\mu \nu} + u^{\mu} q^{\nu} + u^{\nu} q^{\mu} + t^{\mu \nu}
\,,
  \end{aligned}
\end{equation}
where first order derivative dissipative terms for a Bargmann fluid are $\{\eta^\mu,q^\mu,t^{\mu \nu} \}$, which are orthogonal to $\bar\tau_\mu$. These objects can be related to first order derivative terms $\{Q^\mu, j^\mu, t^{\mu \nu} \}$ in the Lorentzian fluid via 
\begin{equation}\label{eq:RelateLorentzian-Bargmann}
  Q^\mu = \frac{1}{c} q^\mu + \frac{1}{c^3} \left( \eta^\mu +... \right) + \CO(c^{-5})\, ,\qquad mj^\mu = \frac{1}{c} q^\mu +\frac{1}{c^3}(...) +\CO(c^{-5})\,,
\end{equation}
where the ellipsis denote terms that do not enter into the constitutive relations. Note that the leading order in $1/c$ of $Q^\mu$ and $mj^\mu$ are equal, due to the Milne boost Ward identity. The field redefinition can then be used to remove either $\eta^\mu$ or $q^\mu$, but not both. This is the analogous to the Eckart and Landau frame in the Lorentzian case. More precisely, following \cite{Jensen:2014ama}, the Landau frame is defined for a theory where the Milne invariant energy density $\tilde \CE^\mu$ receives no derivative correction, i.e., $\tilde \CE = \tilde \CE u^\mu$ and the Eckart frame is defined by demanding the analogous condition for the $U(1)$ current density, i.e., $J^\mu  = \CT^{\mu \nu} \bar\tau_\nu=  n u^\mu$. These conditions are reflected in terms of the constitutive relations \eqref{eq:Milne-invariant-constitutiveReln} as 
\begin{equation}
  \begin{aligned}
\text{(Bargmann) Landau frame :}& \qquad \eta^\mu = 0 \, ,\quad q^\mu \ne 0\,, \\
\text{(Bargmann) Eckart frame :}& \qquad \eta^\mu \ne 0 \, ,\quad q^\mu = 0\,.
  \end{aligned}
\end{equation}
How are these frame choices related to the Eckart and Landau frame in the Lorentzian fluid? Let us first consider the Lorentzian Landau frame where $j^\mu_\text{Lorentz}\ne 0 $. One finds that this gives the non-relativistic theory where $q^\mu = 0$ \cite{Jensen:2014wha}. Similarly, for the Lorentzian Eckart frame, where $J^\mu_\text{Lorentz} =\tilde n U^\mu$, we find that the $U(1)$ current also receives no derivative corrections and thus its $c\to \infty$ limit ends up in the (Bargmann) Eckart frame as well. To summarise this:\footnote{If one starts from a Lorentz fluid in a general frame, one can obtain a Bargmann fluid in Landau frame.}
\begin{equation}
  \left. \begin{matrix}
   \text{(Lorentzian)}\\
  \text{Landau frame :} \;\; Q^\mu =0\, ,j^\mu \ne 0 \\
  \text{Eckart frame :} \;\; Q^\mu \ne0\, ,j^\mu = 0 \\
  \end{matrix}
\qquad \right \}\Longrightarrow\qquad 
\begin{matrix}
  \text{(Bargmann)}\\
  \text{Eckart frame :} \;\; q^\mu =0\, ,\eta^\mu \ne 0 \\
  \\
  \end{matrix}\nonumber
\end{equation}
Essentially, the frame choices in the Lorentzian fluid do not necessarily result in the same frame choice in the Bargmann fluid, once the limit $c\to \infty$ is taken. The actions of picking a frame choice and taking the $c\to \infty$ limit do not commute! 

Working with these Milne invariant quantities is convenient. However, their interpretation is slightly different from the stress-energy tensor $T^{\mu}_{\;\;\,\nu}$ in Eq.~\eqref{eq:WardIdentity}, as the latter does transform under the Milne boost. Moreover, the Milne invariant energy density current $\tilde\CE^\mu$, is not the total energy density of the system and the resulting equations of motion are different from the Ward identity presented in the introduction.\footnote{See e.g. Section 2 of \cite{Jensen:2014ama} for a derivation.} In order to unify notation, we will work with the non-Milne boost invariant quantities, which in flat space, can be obtained via 
\begin{equation}
\begin{aligned}
  T^{0}_{\;\;\, 0} &= \tilde\CE^{0} + \left( u_\nu -\frac{1}{2}\bar\tau_\nu u^2 \right)\CT^{0 \nu}\, ,\\
  T^{i}_{\;\;\, 0} &=\tilde\CE^{i} + \left( u_\nu -\frac{1}{2}\bar\tau_\nu u^2 \right)\CT^{i \nu}\, ,\\
  T^{0}_{\;\;\,i} &=P_i\,, \qquad T^{i}_{\;\;\,j} = \CT^{i \mu}h_{\mu j}\,, \\
\end{aligned}
\end{equation}
and $J^\mu = (J^0,J^i)$. These quantities obey the Ward identity in Eq.~\eqref{eq:WardIdentity}, as shown in \cite{Jensen:2014ama}. Their explicit constitutive relation is given in Section \ref{bargmann_const} and the linear stability analysis can be found in Section \ref{sec:bargmann}. 
%
\subsubsection{Carrollian fluid}\label{sec:car_intro}
The Carroll fluid is obtained as a $c\to0$ contraction of the Poincar\'{e} group \cite{Ciambelli:2018xat,Ciambelli:2018wre,deBoer:2017ing}. The constituitive relations are obtained directly by taking this limit, akin to the massless Galilean case but in contrast to the Bargmann case. This is due to the absence of a central extension for the Carroll algebra in general, see e.g. \cite{Bergshoeff:2017btm,Hartong:2015xda}.\footnote{In some sense the Carroll algebra is already extended since one has $[Q^{i},T_{j}]=\delta^{i}_{\;\;j}H$, where $H$ is the Hamiltonian.}

In order to take the $c\to0$ limit appropriately, one has to take the fluid velocity $v^{i}$ faster to zero that $c$ in this limit, in order to avoid branch cuts. We execute this by introducing inverse velocity $w^{i}$ and equating $v^{i}=c^{2}w^{i}$ \cite{levy1965nouvelle}, which we write more covariantly as $w^{\mu}=P^{\mu}_{\;\;\,i}w^{i}$ and $\omega_{\nu}=h_{\nu i}w^{i}$.
 
The constitutive relations can be obtained from Lorentzian fluid constitutive relations by taking the limit $c\to 0$,
\begin{equation}\label{eq:cToZeroLimit}
	\begin{aligned}
	U^\mu 
	&= 
	\tau^\mu + c^2 \left( w^\mu + \frac{1}{2}\tau^\mu w^2 \right) + \CO(c^3)\, ,\quad
  	U_\nu 
	= 
	c^2 \left( \bar\tau_\nu + w_\nu \right)+ \CO(1)\, ,
  	\\
	\eta_{\mu \nu} 
	&= 
	h_{\mu \nu} + \CO(c^2)\, , 
	\qquad 
	\eta^{\mu \nu} 
	= 
	-\frac{1}{c^2}\tau^\mu \tau^\nu + \CO(1) \, ,
	\\
	\Delta^{\mu}_{\;\;\,\nu} 
	&= 
	\delta^{\mu}_{\;\; \nu} - \tau^\mu \bar\tau_\nu + \tau^\mu w_\nu\,,
	\quad
	\Delta^{\mu \nu} 
	= 
	h^{\mu \nu} +w^2 \tau^\mu \tau^\nu + \tau^{\mu}w^{\nu}+\tau^{\nu}w^{\mu} + \CO(c) 
	\,,
	\end{aligned}
\end{equation}
while the transport coefficient is scaled as 
\begin{equation}
  \eta \to \eta/c^2 \, ,\qquad \zeta \to \zeta/c^2 \,, \qquad \kappa\to c^2 \kappa \,.
  \end{equation}
Note that in this limit, $q^\mu U_\nu \sim \CO(c^2)$ while $q_\nu U^\mu \sim \CO(1)$. 

There are several peculiar features of the Carrollian fluid. Firstly, while the fluid velocity $v^i \to 0$ in the $c\to 0$ limit, the momentum density flux $P_i = T^0_{\;\;\, i}$ is non-zero when $w^i \ne 0$. In other words, momentum density flux is not related to the fluid velocity. This is in conjunction with the result in \cite{Bergshoeff:2014jla}, which claims that a Carollian particle cannot move. Their momentum density and $U(1)$ current expectation value can, nevertheless, be nonzero and can be transported by a diffusion process, as we will show in Section \ref{sec:carroll}.

Secondly, the Ward identity of the Carroll boost, $T^{i}_{\;\;\,0}=0$, puts the energy density flux to zero by construction. This makes the notion of the Landau frame in \eqref{eq:landau_frame} rather obscure. Nevertheless, as one obtains explicit constitutive relations for the Carrollian fluid via the Lorentz fluid, some of the former's properties are inherited from the latter. For example, the Lorentzian Eckart frame condition $j^\mu = 0$ persists as one takes $c\to 0$ limit, just as what one would expect. We note, however, that we cannot relate these resulting frames to one-another using a shift in $\mu$, $T$ and $w^{\mu}$, which we will soon show in Section \ref{carroll_const}. This is curious, but does not change the fact that we have a self consistent Carroll fluid. We will also present the explicit constitutive relations in Section \ref{carroll_const} and, in Section \ref{sec:carroll}, we present a study of the stability of the resulting fluids.

%
\subsection{Explicit constituitive relations}
From a perspective of a unified framework and in the spirit of \cite{deBoer:2017abi}, we present a translationally and spatial rotationally invariant fluid, including a $U(1)$ symmetry as well, from which we can reproduce the constitutive relations for Bargmann and Carroll, by invoking the relevant (boost) Ward identities. We can also reproduce a specific fluid without boost invariance.
The constitutive relations of such a unified fluid can be given by
\begin{equation}\begin{aligned}\label{eq:aristotle_const}
	T^{\mu}_{\;\;\,\nu}
	=&
	-
	\tilde{E}_{1} u^{\mu}\overline{\tau}_{\nu}
	-
	\tilde{E}_{2} \tau^{\mu}\overline{\tau}_{\nu}
	+
	\mathcal{P}_{1}\Pi^{\mu}_{\;\;\,\nu}
	+
	\mathcal{P}_{2}
	\left(
		\delta^{\mu}_{\;\;\nu}
		-
		\tau^{\mu}\overline{\tau}_{\mu}
	\right)
	+
	\mathcal{M}_{1} u^{\mu}
	\Gamma_{\nu}
	+
	\mathcal{M}_{2} \tau^{\mu}
	\omega_{\nu}
	\\&
	+
	i^{\mu}u_{\nu}
	-
	q^{\mu}\overline{\tau}_{\nu}
	+
	u^{\mu}\overline{q}_{1\nu}
	+
	\tau^{\mu}\overline{q}_{2\nu}
	+
	t^{\mu}_{\;\;\,\nu}
	\,,
	\\
	J^{\mu}
	=&\;
	\mathcal{N}_{1} u^{\mu}
	+
	\mathcal{N}_{2} \tau^{\mu}
	+
	j^{\mu}
	\,,
\end{aligned}\end{equation}
where $\Gamma_{\nu}:=u_{\nu}-v^{2}\overline{\tau}_{\nu}$ and $t^{\mu}_{\;\;\nu}$ is traceless. In the current context we define $\Pi^{\mu}_{\;\;\,\nu}=\delta^{\mu}_{\;\;\nu}-u^{\mu}\overline{\tau}_{\nu}$ and $u_{\mu}=h_{\mu\nu}u^{\nu}$. The introduced scalars, vectors and tensor above are defined in the same vain as the relativistic case, see text below Eq. \eqref{eq:general-Lorentzian-constitutiveReln}, with respect to taking into account the gradient expansion.
From the thermodynamics point of view, we have a slightly more general first law and Euler relation than in the relativistic case:
\begin{equation}\label{eq:aristotle_thermo}
	d\tilde{\mathcal{E}}
	=
	Tds
	+
	\mu dn
	-
	\frac{1}{2}\rho dv^{2}
	\,,
	\quad
	\tilde{\mathcal{E}}
	+
	P
	=
	T
	s
	+
	\mu
	n
	\,,
\end{equation}
where $\rho$ is generalized mass density \cite{deBoer:2017ing}. Since there is a priori no boost symmetry, this generalized mass density can be thought of as a chemical potential.

Finally we comment on the entropy current constraint \cite{LLfluid}, as presented in Eq.~\eqref{eq:entropyProd}. We can typically express the entropy current divergence $\partial_{\mu} s^\mu$ in terms of the constitutive relations. This especially means that $ \partial_{\mu} s^\mu$ is expressed in terms of transport coefficients such as viscosity. Requiring $\partial_{\mu}s^{\mu}\geq0$ for all possible fluid configurations (not just on-shell), implies conditions on the transport coefficients. This condition was explicitly checked for all fluids used in this note, except for the Carrollian fluid. However, since the Carrollian fluid is a limit of the Lorentzian fluid, for which we did apply this entropy current condition, it is reasonable to assert that the Carrollian fluid satisfies the entropy current condition.
\subsubsection{Bargmann constitutive relations}\label{bargmann_const}
As a result of Milne or Galilean boost invariance, we obtain the Ward identity $T^{0}_{\;\;\,i}=m J^{i}$. Here $m$ is the particle mass, since typically $J^{\mu}$ is considered to be related to particle number conservation in this setup. As a result: $\rho=m n$. 

Using the above relation between $\rho$ and $n$, we can rewrite Eq. \eqref{eq:aristotle_thermo} to be
\begin{equation}
	d\hat{\mathcal{E}}
	=
	Tds
	+
	\hat{\mu} dn
	\,,
	\quad
	\hat{\mathcal{E}}
	+
	P
	=
	T
	s
	+
	\hat{\mu}
	n
	\,,
\end{equation}
where we redefined
\begin{equation}
	\hat{\mathcal{E}}
	=
	\tilde{\mathcal{E}}+\frac{1}{2}mnv^{2}
	\,,
	\quad
	\hat{\mu}
	=
	\mu
	+
	\frac{1}{2}mv^{2}
	\,.
\end{equation}
Let us now give the constitutive relations of this fluid, using the framework presented by the more general fluid in Eq. \eqref{eq:aristotle_const}. For a Bargmann fluid, in the here considered frames, one finds \cite{Jensen:2014ama}
\begin{equation}
	\tilde{E}_{1}
	=
	\tilde{\mathcal{E}}
	\,,
	\quad
	\mathcal{P}_{1}
	=
	P-\zeta\partial_{\lambda}u^{\lambda}
	\,,
	\quad
	\mathcal{M}_{1}
	=
	m
	\mathcal{N}_{1}
	=
	n
	\,,
	\quad
	t^{\mu}_{\;\;\,\nu}
	=
	\eta
	\sigma^{\mu}_{\;\;\,\nu}
	\,,
\end{equation}
where $\sigma^{\mu}_{\;\;\,\nu}:=\lim_{c\to\infty}\Sigma^{\mu}_{\;\;\,\nu}=h^{\mu\alpha}\left(\overline{\tau}_{\nu}h^{\beta\rho}u_{\rho}-P^{\beta}_{\;\;\nu}\right)\left(\partial_{\alpha}u_{\beta}+\partial_{\beta}u^{\alpha}-\frac{2}{d}h_{\alpha\beta}\partial_{\lambda}u^{\lambda}\right)$ and $\tilde{\mathcal{E}}_{2}=\mathcal{P}_{2}=\mathcal{M}_{2}=\mathcal{N}_{2}=q_{2\nu}=0$.
In the \textbf{Eckart frame} one finds
\begin{equation}
	i^{\mu}
	=
	j^{\mu}
	=
	0
	\,,
	\quad
	q^{\mu}
	=
	-\kappa h^{\mu\alpha}\partial_{\alpha}T
	\,,
	\quad
	\overline{q}_{1\nu}
	=
	0
	\,.
\end{equation}
Similar to the Lorentz case, the Eckart frame is defined as having vanishing particle density flux. For the \textbf{Landau frame} it is found that
\begin{equation}
	i^{\mu}
	=
	j^{\mu}
	=
	V^{\mu}
	\,,\quad
	q^{\mu}
	=
	-\frac{1}{2}u^{2}V^{\mu}
	\,,
	\quad
	\overline{q}_{1\nu}
	=
	\left(
		h_{\nu\lambda}
		-
		\overline{\tau}_{\nu}u_{\lambda}
	\right)V^{\lambda}
	\,,
\end{equation}
with $V^{\mu}=-\sigma u^{\nu}\partial_{\nu}u^{\mu}-\sigma T P^{\mu}_{\;\;\lambda}\partial^{\lambda}\frac{\hat{\mu}}{T}$.
The Landau frame is defined as vanishing Galilean \textit{boost invariant} energy density flux \cite{Jensen:2014ama}. Furthermore, $\kappa=\sigma(\hat{\mathcal{E}}+P)^{2}/(n^{2}T)$. 
The \textbf{general frame} is given by
\begin{equation}
	i^{\mu}
	=
	j^{\mu}
	=
	V^{\mu}
	\,,\quad
	q^{\mu}
	=
	-\frac{1}{2}u^{2}V^{\mu}
	-
	\kappa h^{\mu\alpha}\partial_{\alpha}T
	\,,
	\quad
	\overline{q}_{1\nu}
	=
	\left(
		h_{\nu\lambda}
		-
		\overline{\tau}_{\nu}u_{\lambda}
	\right)V^{\lambda}
	\,.
\end{equation}
In this frame, $\kappa$ and $\sigma$ are independent. In all the here presented cases $\kappa$, $\sigma$, $\zeta$ and $\eta$ are positive, due to explicit check of the entropy current.
\subsubsection{Carroll constitutive relations}\label{carroll_const}
A Carrollian fluid has to obey the Ward identity $T^{i}_{\;\;0}=0$. In other words: the energy density flux of a Carrollian fluid always vanishes. The thermodynamics are given by
\begin{equation}
	d\tilde{\mathcal{E}}
	=
	Tds
	+
	\mu dn
	\,,
	\quad
	\tilde{\mathcal{E}}
	+
	P
	=
	T
	s
	+
	\mu
	n
	\,,
\end{equation}
since $v^{2}\to0$. The constitutive relations can be fit into the framework in Eq.~\eqref{eq:aristotle_const} using
\begin{equation}\begin{aligned}
	\tilde{E}_{2}
	=&\;
	\tilde{\mathcal{E}}
	\,,
	\quad
	\mathcal{P}_{2}=P-\zeta \partial_{\rho} \left(w^\rho + \frac{1}{2}  w^2 \tau^\rho  \right)
	\,,\quad
	\mathcal{M}_{2}
	=
	\tilde{\mathcal{E}}+P
	\,,
	\quad
	\mathcal{N}_{2}=n
	\,,
	\quad
	i^{\mu}=q^{\mu}
	=
	0
	\,,
	\\
	t_{\;\;\,\nu}^{\mu} =& -\eta 
	\left[
		\tilde{\Pi}^{\mu\rho}
	  	\partial_{\rho} w_\nu 
		+ 
		h^{\mu \rho} \tilde{\Pi}^{\sigma}_{\;\;\,\nu}\d_{\sigma} w_\rho 
		+
		\tau^\mu w^\rho \tilde{\Pi}^{\sigma}_{\;\;\,\nu}\d_{\sigma} w_\rho 
		- 
		\frac{2}{d} \tilde{\Pi}^{\mu \rho} h_{\rho \nu} 
		\d_\lambda \left(w^\lambda
			+ 
			\frac{1}{2}w^2 \tau^\lambda  
		\right)   
	\right]\,,
\end{aligned}\end{equation}
with $\tilde{\mathcal{E}}_{1}=\mathcal{P}_{1}=\mathcal{M}_{1}=\mathcal{N}_{1}=\overline{q}_{1\nu}=0$, $\tilde{\Pi}^{\mu}_{\;\;\,\nu}=\lim_{c\to0}\Delta^{\mu}_{\;\;\,\nu}=\delta^{\mu}_{\;\;\,\nu}-\tau^{\mu}\overline{\tau}_{\nu}+\tau^{\mu}w_{\nu}$ and $\tilde{\Pi}^{\mu\nu}=\lim_{c\to0}\Delta^{\mu\nu}=h^{\mu \nu} +w^2 \tau^\mu \tau^\nu + \tau^{\mu}w^{\nu}+\tau^{\nu}w^{\mu}$. One finds that these Carroll projectors are transverse to $\tau^{\mu}$ and $\tau_{\nu}-w_{\nu}$. The results for these constitutive relations are obtained by taking $c\to0$, such that the result automatically satisfies the entropy production constraint.

Starting from the \textbf{Landau frame in the Lorentzian fluid}, one can obtain the Carrollian constitutive relations using \eqref{eq:cToZeroLimit}. The resulting first derivative terms can be written as
\begin{equation}
	\overline{q}_{2\nu}
	=
	0
	\,,
	\quad
	j^\mu 
	= 
	- \sigma \tilde{\Pi}^{\mu\nu}\d_\nu \frac{\mu}{T}
	\,.
\end{equation}
If, instead, we take $c\to 0$ limit, starting from \textbf{Eckart frame of Lorentzian fluid}, we find that
\begin{equation}
	 \overline{q}_{2\nu} = -\kappa \left(\d_\nu T - \bar\tau_\nu \tau^\alpha \d_\alpha 	T +T \tau^\lambda \d_\lambda w_\nu + w_\nu \tau^\alpha \d_\alpha T  \right)
	 \,,
	 \quad
	 j^{\mu}
	 =
	 0
	 \,.
\end{equation}
We retain the relation $\kappa=\sigma(\tilde{\mathcal{E}}+P)^{2}/(n^{2}T)$. In the case where the limit $c\to 0$ is taken from the Lorentzian fluid's \textbf{general frame}, we have
\begin{equation}
	\overline{q}_{2\nu} = -\kappa \left(\d_\nu T - \bar\tau_\nu \tau^\alpha \d_\alpha 	T +T \tau^\lambda \d_\lambda w_\nu + w_\nu \tau^\alpha \d_\alpha T  \right)
	\,,
	\quad
	j^\mu 
	= 
	- \sigma \tilde{\Pi}^{\mu\nu}\d_\nu \frac{\mu}{T}
	 \,.
\end{equation}
Here $\kappa$ and $\sigma$ are in principle independent. In all cases here, $\kappa$, $\sigma$, $\eta$ and $\zeta$ are non-negative due to the positivity of $\d_\mu s^\mu$.

We want to focus some attention to the choice of frames in the $c\to0$ contraction. In order to call the Landau frame and Eckart frame, as presented above, true frame choices of a Carroll fluid, these frames should be connected by a frame transformation, which can be induced by invoking \cite{Kovtun:2012rj}
\begin{equation}\label{eq:fields}
	T\to T+\eth T
	\,,
	\quad
	\mu
	\to
	\mu+\eth\mu 
	\,,
	\quad
	w^{\mu}
	\to
	w^{i}	
	+
	\eth w^{i}
	\,,
\end{equation}
where terms accompanied by $\eth$ denote those terms to be of first order in derivatives. Focussing on the influence of this transformation on $J^{\mu}$, we see
\begin{equation}
  J^\mu(T+ \eth T,\mu + \eth \mu + w^i + \eth w^i) = (n + \eth n) \tau^\mu +j^\mu + \CO(\d^2)
\end{equation}
where 
$\mathcal{N}_{2}(T+\eth T,\mu+\eth\mu, w^{i}+\eth w^{i})=n+\eth n(\eth T,\eth\mu,\eth w^{i}) + \CO(\d^2)$. 
Thus, using a transformation as in \eqref{eq:fields}, which enables frame transformations, one is only able to fix the zeroth component $j^0$ of $j^{\mu}$. In other words, the Landau frame and Eckart frame presented above cannot be related by a frame choice, since this would require changing the $i$ component of $j^{i}$. This issue ought to be sorted out by constructing a Carroll fluid directly, rather than using reductions from a Lorentzian fluid. However, that is beyond the scope of this paper. 
\subsubsection{A more general fluid: relaxing the boost constraint}\label{sec:general}
In \cite{deBoer:2017abi}, a static fluid without (necessarily) imposing boost symmetry, but with translational and rotational invariance, was studied. This was done in a linearised setting, where the fluid background velocity was put to zero, i.e. $u^{\mu}=\tau^{\mu}+\delta u^{\mu}$. In order to fit the results of \cite{deBoer:2017abi} in the currently used framework, we restricted ourself to the case where $\tilde{E}_{2}=0$, $\mathcal{P}_{2}=0$, $\mathcal{M}_{2}=0$, $\mathcal{N}_{2}=0$, $\overline{q}_{2\nu}=0$ in \eqref{eq:aristotle_const}. One obtains the following linearised constitutive relations
\begin{equation}\begin{aligned}\label{eq:emt_gen_2}
	T^{\mu}_{\;\;\,\nu}
	=&
	-
	\left(
		\tilde{\mathcal{E}}+\delta \tilde{E}_{1}
	\right)\tau^{\mu}\overline{\tau}_{\nu}
	+
	\left(
		P+\delta\mathcal{P}_{1}
	\right)P^{\mu}_{\;\;\,\nu}
	-
	\left(
		\tilde{\mathcal{E}}
		+
		P
	\right)\delta u^{\mu}\overline{\tau}_{\nu}
	+
	\rho \tau^{\mu}\delta u_{\nu}
	+
	\tau^{\mu}
	\delta\overline{q}_{1\nu}
	-
	\delta q^{\mu}_{1}\overline{\tau}_{\nu}
	\,,
	\\
	J^{\mu}
	=&
	n\tau^{\mu}
	+
	n\delta u^{\mu}
	+
	\delta \mathcal{N}_{1}\tau^{\mu}
	+
	\delta j^{\mu}
	\,,
\end{aligned}\end{equation}
where $\delta$ denotes that an object is exactly of order linear in derivatives (all others are constants). It turns out that $\delta j^{\mu}=0$ is equivalent to the Landau frame, whereas $\delta q^{\mu}=0$ is equivalent to Eckart frame. In both frames we have
\begin{equation}
	\tilde{E}_{1}
	=
	\tilde{\mathcal{E}}
	\,,
	\quad
	\mathcal{P}_{1}
	=
	P-\zeta\partial_{\lambda}u^{\lambda}
	\,,
	\quad
	\mathcal{M}_{1}
	=
	\rho
	\,,
	\quad
	\mathcal{N}_{1}
	=
	n
	\,,
\end{equation}
\begin{equation}
	\overline{q}_{1\nu}=-\overline{\pi}\partial_{t}\delta u_{\nu}+\overline{\alpha}T
 P^{\alpha}_{\;\;\,\nu}\partial_{\alpha}\delta\frac{\mu}{T}
	\,,
	\quad
	\delta\tilde{E}_{1}=\delta \mathcal{N}_{1}=0
	\,,
	\quad
	\delta\mathcal{P}_{1}=-\overline{\zeta}\partial_{\kappa}\delta u^{\kappa}
	\,,
	\quad
	\delta t^{\mu}_{L\,\nu}=\eta \delta \sigma^{\mu}_{\;\;\,\nu}	
	\,.
\end{equation}
Now one has the relations
\begin{equation}
	\zeta
	=
	\overline{\zeta}
	+
	f(\tilde{a}_{T},\tilde{a}_{\frac{\mu}{T}})
	\,,
	\quad
	\pi 
	=
	\overline{\pi}
	-
	\tilde{a}_{T}
	\,,
	\quad
	\alpha
	=
	\overline{\alpha}
	-
	\tilde{a}_{\frac{\mu}{T}}
	\,,
\end{equation}
where $\tilde{a}_{T}$ and $\tilde{a}_{\frac{\mu}{T}}$ are some functions of $a(T,\frac{\mu}{T})$, which is a non-dissipative transport coefficient. That means that $a(T,\frac{\mu}{T})$, and thus also $\tilde{a}_{T}$ and $\tilde{a}_{\frac{\mu}{T}}$, does not appear in the entropy current and therefore is not restricted to any specific range of values. Here $f(\tilde{a}_{T},\tilde{a}_{\frac{\mu}{T}})$ is some function of the two arguments and various thermodynamic quantities. This object will not play any role in our further analysis.

Furthermore we have dissipative transport coefficients: $\overline{\zeta}\geq 0$, $\eta\geq0$, $\overline{\pi}\geq0$, $\sigma \geq0$, $\overline{\alpha}^{2}\leq \overline{\pi}\sigma$. These bounds on the transport coefficients are found by examining the entropy current as done in \cite{deBoer:2017abi}. In the \textbf{Landau frame} ($\delta q^{\mu}
	=
	0$) one finds
\begin{equation}\label{eq:genlau}
	\delta j^{\mu}=
	\overline{\alpha} \partial_{t}\delta u^{\mu}
	-
	\sigma T  h^{\mu\nu}\partial_{\nu}\delta\frac{\mu}{T}
	\,,
	\quad
	\delta\overline{q}_{1\nu}
	=
	-\overline{\pi}\partial_{t}\delta u_{\nu}+\overline{\alpha}T
 P^{\alpha}_{\;\;\,\nu}\partial_{\alpha}\delta\frac{\mu}{T}
	\,.
\end{equation}
We can construct the Eckart frame by considering a transformation of variables from the results of the Landau frame, of which the technical details can be found in Appendix \ref{appendix1}. For the \textbf{Eckart frame} ($\delta j^{\mu}=0$) we find
\begin{equation}\label{eq:geneck}
	\delta \overline{q}_{1\,\nu}
	\!=\!
	-
	\!
	\left(
		\overline{\pi}
		+
		\overline{\alpha}\frac{\rho}{n}
	\right)
	\!
	\partial_{t}\delta u_{\nu}
	+
	\!
	\left(
		\overline{\alpha}
		+
		\sigma\frac{\rho}{n} 
	\right)
	\!
	T
 P^{\alpha}_{\;\;\,\nu}\partial_{\alpha}\delta\frac{\mu}{T}
	\,,
	\;
	\delta q^{\mu}_{1}
	\!=\!
	\frac{\tilde{\mathcal{E}}+P}{n} 
	\!
	\left(
		\overline{\alpha} \partial_{t}\delta u^{\mu}
		-
		\sigma T  h^{\mu\nu}\partial_{\nu}\delta\frac{\mu}{T}
	\right)
	\!
	\,.
\end{equation}
In Section \ref{generalfluid}, we consider the stability conditions for this fluid. Furthermore, we can reproduce the Bargmann and Lorentz results directly, by applying the relevant Ward identities (Carroll can be obtained by employing the familiar $c\to\infty$ contraction). For the different boost cases, the frame choices coincide with what is done for the more general fluid, due to being in a linearised and static regime.
\section{Instabilities of fluids with Galilean boost symmetries}\label{sec:bargmann}
In this section, we study the spectrum of the fluid with Bargmann symmetry. We will focus on a fluid that lives in flat $3+1$ spacetime dimensions. As discussed in Section \ref{sec:technicalOverview}, choosing the Eckart or Landau frame in the Lorentzian fluid and taking $c\to \infty$, results in different constitutive relations from taking $c\to \infty$ first and then choosing the frame in the Bargmann fluid. To keep the physical picture consistent, we shall refer to the Eckart frame in this section as the theory with $J^\mu = \tilde n u^\mu$, without a derivative correction \textit{after taking} $c\to \infty$ limit. Similarly, the Landau frame will refer to a theory with Galilean boost invariance where the Milne invariant energy density $\tilde \CE^\mu $ receives no derivative corrections. Their relations with frame choices in the Lorentzian fluid are discussed in Section \ref{sec:technicalOverview-Bargmann}. 

It turns out that the Landau frame in this theory is unstable even when the fluid is static and homogeneous. The Eckart frame, on the other hand, is stable and remains so when the fluid acquires a finite momentum. In hindsight, the stability of the Bargmann fluid's Eckart frame is not surprising since it reproduces precisely the Navier-Stokes equations, as is well known. What is more interesting, is that a general frame of non-relativistic (Bargmann) fluid, like the relativistic fluid, is unstable. It turns out that the Eckart frame is a special case. 

\subsection{Landau frame}
We follow the standard procedure of linear perturbation in e.g. \cite{Martin:1963}, namely 
\begin{equation}
  \lambda(t,x^i) \sim \lambda + \delta \lambda \, e^{-i \omega t + i k_i x^i}\,,\qquad \lambda = \{T, \hat \mu,  v^i \}
  \,,
\end{equation}
where the other thermodynamic quantities, such as $\hat \CE,P,n$ depend only on $T$ and $\hat{\mu}$. We also set $m=1$ for simplicity. Let us first focus on the transverse fluctuations, consisting of $\delta v_\perp^{i}$ which is a fluid velocity in the direction perpendicular to wave vector $k_i$. The spectrum for transverse momentum of the Bargmann fluid in this frame is governed by the following polynomial 
\begin{equation}
0 = \omega \left(-i n + \omega\sigma  \right)   + i k_i v^i \left(n + 2i\omega\sigma  \right) + k^2 \left( \eta + v^2 \sigma \right)\,,
\end{equation}
where $v^i$ and $v^2 = v_i v^i$ correspond to the background velocity of the fluid.
Solving for $\omega(k)$, we find that it consists of two modes 
\begin{equation}\label{eq:Landau-Bargman-transverse-spectrum}
  \omega_\perp^- = k_i v^i - i\left( \frac{\eta}{n} \right)k^2 + \CO(k^3)\,,\qquad 
  \omega^+_\perp = k_i v^i + i\left( \frac{n}{\sigma}\right) + i\left( \frac{\eta}{n} \right)k^2+ \CO(k^3)
  \,.
\end{equation}
The second pole, $\omega^+_\perp$, is the unphysical unstable mode, analogous to the one in Lorentzian fluid's Eckart frame, as pointed out in \cite{Hiscock:1985zz}. Unlike the Landau frame in the Lorentzian fluid, here $v^i \to 0$ is not a singular limit and we can simply take $v^i \to 0$ in Eq. \eqref{eq:Landau-Bargman-transverse-spectrum}, in order to obtain the spectrum for the fluid at rest. There, we see that the positive imaginary part still persists in $\omega^+_\perp$, which indicates that the theory is unstable even around a static configuration. 

Let us also look at the longitudinal fluctuations consisting of $\{\delta T, \delta \hat{\mu}, \delta v^i_\parallel \}$. For simplicity let us choose $k^i = (k,0,0)$ and the background velocity $v^i=(v,0,0)$. The spectrum of modes in the longitudinal fluctuations is governed by the quartic equation in $\omega$. To see this, one can write down the Fourier transformed equations of motion as 
\begin{equation}\label{eq:eom-metM-Bargman}
  \mathbf{M}(\omega,k)
  \begin{pmatrix}
\delta T \\
\delta \hat{\mu} \\
\delta v_x 
  \end{pmatrix}\,  = 0\,,
\end{equation}
where the matrix $\mathbf{M}(\omega,k) = -i\omega \mathbf{A}(\omega,k) +ik \mathbf{B}(\omega,k)$ can be written as 
\begin{equation}
\begin{aligned}
\mathbf{A} &=
\begin{pmatrix}
\alpha_1 + i \left(\frac{\hat\mu \sigma}{T}  \right) kv  & \alpha_2 + i \sigma kv & \alpha_3 v \\
\chi_{21}+ i\left( \frac{\hat \mu \sigma}{T}\right) k & \chi_{22} v- i k \sigma & \alpha_3 \\
\chi_{21} & \chi_{22} & 0
\end{pmatrix} \,,
\end{aligned}
\label{eq:metA-BargmanSound}
\end{equation}
where functions $\alpha_1 =T \chi_{11} + \chi_{12} (\hat\mu + v^2/2)$, $\alpha_2 = T\chi_{12} + \chi_{22}(\hat\mu + v^2/2)$ and $\alpha_3 = n + i \sigma (\omega-kv)$. The matrix $\mathbf{B}$ can be written as 
\begin{equation}
\begin{aligned}
\mathbf{B} &= 
\begin{pmatrix}
\alpha_1v + i\left(\frac{\hat\mu \sigma}{2T} kv^2  \right) & \alpha_2 v + i\frac{\sigma}{2} kv^2 & \beta \\
s + \chi_{21}v^2 +2i\left( \frac{\hat\mu \sigma}{T} \right)kv & n + \chi_{22}v^2 -2i \sigma kv & 2 \alpha_3v -i \gamma k\\
\chi_{21}v + i\left( \frac{\hat\mu \sigma}{T} \right)k & \chi_{22}v - i \sigma k & n + i \sigma(\omega - kv)
\end{pmatrix}
\,.
\end{aligned}
\label{eq:metB-BargmanSound}
\end{equation}
Here $\beta = \hat{\CE} + P + \frac{3}{2} nv^2 - i \gamma kv - i \sigma v^2 (\omega -kv)$, $\gamma = \zeta + 4\eta/3$. The susceptibility matrix is defined via 
$\chi_{11} = (\d s/\d T)_{\hat\mu}$, $\chi_{12} = (\d s/\d \hat\mu)_T$, $\chi_{21} = (\d n/\d T)_{\hat\mu}$ and $\chi_{22} = (\d n/\d \hat\mu)_T$.

The spectrum can be determined by taking $\det[\mathbf{M} ]= 0$\,, which results in a quartic equation in $\omega$ due to the $\omega$ dependence in $\mathbf{A}$. 
This implies that there are four solutions to this system. There are two complex solutions corresponding to the sound modes and one purely imaginary solution corresponding to the thermoelectric diffusion (see e.g. \cite{Kovtun:2012rj,Davison:2015taa} for discussion). The other mode is gapped and contains a positive imaginary part, its spectrum can be written as 
\begin{equation}\label{eq:Bargman-sound-unstable}
  \omega^+_\parallel = i\left(\frac{\rho}{\sigma}\right) 
  + k v + i \left( \frac{\gamma}{n} + \frac{ \sigma\chi_{22}(\hat \CE+P)(\hat\CE+P+2 n v^2)}{n^2T^2 \det\chi} \right)k^2+\CO(k^3)
  \,,
\end{equation}
which also leads to a linear instability.

\subsection{Eckart frame}

This constitutive relation is the one used in the majority of \cite{LLfluid} and, in the limit where $ n \gg T$, is nothing but the standard Navier-Stokes equations found in the literature. This is a very successful effective description of everyday phenomena and exhibits no artificial instabilities of the kind discussed in this work. 

In order to explicitly see the linear stability, we consider the same setup as in the Landau frame. First, let us look at the perturbation around the fluid with a constant velocity $v^i \ne 0$. The equation of motion and the spectrum can be found in e.g. \cite{LLfluid} and we will not repeat it here. It turns out that the spectrum in the transverse channel is linear in $\omega$ unlike in the Landau frame, and yields the standard diffusion dispersion relation
\begin{equation}
    \omega_\perp = k_i v^i - i\left( \frac{\eta}{n}\right) k^2 + \CO(k^3)
    \,.
\end{equation}
In the longitudinal channel, the equation that governs the spectrum is now a cubic equation due to the absence of a term proportional to $\omega\sigma$ in $\mathbf{A}$. Unlike the Landau frame in \eqref{eq:metA-BargmanSound}, we have: 
\begin{equation}
  \mathbf{A}_\text{Eckart} = \mathbf{A}_\text{Landau}\Big\vert_{\sigma = 0} \,,\qquad \mathbf{B}_\text{Eckart} = \mathbf{B}_\text{Landau}\Big\vert_{\sigma = 0} + 
ik   \begin{pmatrix}
\kappa & 0 & 0 \\
0 & 0 & 0\\
0 & 0 & 0 
  \end{pmatrix} 
  \,,
\end{equation}
The determinant of $3\times 3$ matrix $\mathbf{M} = -i \omega \mathbf{A} + ik \mathbf{B}$ therefore yields a cubic polynomial in $\omega$, indicating that there are only three poles in the correlation functions. These are the sound mode and the thermoelectric diffusion. There are no unstable modes in the spectrum. 
\subsection{General frame}

One can also analyse the spectrum of the theory without choosing either Landau or Eckart frame. The spectrum for transverse fluctuations is identical to those in Landau frame, namely Eq.~\eqref{eq:Landau-Bargman-transverse-spectrum} and exhibit the same instability. Similarly, the equation of motion for the longitudinal fluctuations can be written in the same form as Eq.~\eqref{eq:eom-metM-Bargman}. Writing $\mathbf{M} =-i\omega \mathbf{A} + i k \mathbf{B}$, the matrices $\mathbf{A}, \mathbf{B}$ are 
\begin{equation}
  \mathbf{A} = \mathbf{A}_\text{Landau} \,, \qquad \mathbf{B} = \mathbf{B}_\text{Landau} + i + 
ik   \begin{pmatrix}
\kappa & 0 & 0 \\
0 & 0 & 0\\
0 & 0 & 0 
\end{pmatrix}
\,,
\end{equation}
where $\mathbf{A}_\text{Landau},\mathbf{B}_\text{Landau}$ are taken from \eqref{eq:metA-BargmanSound} and \eqref{eq:metB-BargmanSound}. As in the Landau frame case, the spectrum in the longitudinal channel is governed by the quartic polynomial of $\omega$ and yields the same unstable mode as in \eqref{eq:Bargman-sound-unstable}.

We may also think of the Eckart frame as a limit where $\sigma\to 0$. It turns out that this choice of parameter is a singular limit. This can be easily seen from the spectrum of the transverse channel spectrum in Eq.~\eqref{eq:Landau-Bargman-transverse-spectrum} and the unstable pole in the longitudinal channel Eq.~\eqref{eq:Bargman-sound-unstable} . The unstable pole becomes more unstable as we approach $\sigma \to 0$ limit as $\omega \to +i\infty$ in the complex $\omega-$plane. The situation is opposite to the Lorentzian fluid, where the same scenario occurs as one moves from the general to the Landau frame \cite{Hiscock:1985zz} instead of the Eckart frame.

\newpage
\section{Instabilities of fluids with Carrollian boost}\label{sec:carroll}
We analyse the spectrum of the Carroll fluids given in Section \ref{bargmann_const}, starting from Landau frame followed by Eckart frame and end with the general frame. Recall that these frames were defined via the Lorentzian parent. We do this in 3+1 dimensional flat spacetime. 

To summarise our findings: the Landau frame with $P_i =T^{0}_{\;\;\,i}=0$ is stable, but becomes unstable when $P_i \ne 0$. The Eckart frame is unstable even in the $P_i = 0$ case. This pattern resembles the instabilities of the Lorentzian fluid. We would like to note the absence of any sound mode in the Carrollian fluid. This is due to the fact that conservation of energy density $\d_\mu T^\mu_{\;\; 0} = 0$ implies that $\d_0 \tilde \CE = 0$ and therefore it relates the fluctuation of temperature $\delta T$ and the chemical potential $\delta \mu$. The only non-trivial dynamics of $\delta T$ (or $\delta \mu$ depending on our preferred choice of variables), can be made to decouple from the longitudinal momentum.

We shall elaborate further on the absence of the sound mode. To do this, it is convenient to write down the thermodynamic quantities as functions of energy density $\tilde \CE$ and the $U(1)$ density $n$. The Carrollian Ward identity implies that the energy density fluctuation $\delta \tilde \CE =0$. As a result, the dynamical variables are $\delta n$ and $\delta w^i$ in the longitudinal channel. In the case where the background inverse velocity $w^i =0$, we find the following equations of motion
\begin{equation}
  0 = \left[-i\omega \begin{pmatrix}
-i k \left(\frac{\d T}{\d n}\right)_{\tilde \CE} \kappa  & (\tilde \CE+P + i\omega T\kappa) \\
1 & 0
  \end{pmatrix} + i k 
  \begin{pmatrix}
\left(\frac{\d P}{\d n}\right)_{\tilde \CE} & -ik \left(  \zeta + \frac{4}{3}\eta\right)\\
0 & -ik \frac{\sigma}{T^2} \left(T\frac{\d \mu}{\d n}- \mu\frac{\d T}{\d n}  \right)
  \end{pmatrix}\right] \begin{pmatrix} 
\delta n \\ \delta w^x
  \end{pmatrix}
\end{equation}
We should note the difference between this and the equations of motion for the longitudinal channel in the Bargmann and Lorentzian fluid. Firstly, in the ideal Carrollian fluid (where $\zeta,\eta,\kappa,\sigma$ vanish), the only solution is $\omega = 0$, instead of $\omega = \pm c_s k$ where $c_s$ is the speed of sound. This signifies the fact that when the first order correction is turned on, we find that are no modes for which $\text{Re\,}\omega = 0$. 

\subsection{$c\to 0$ limit of Lorentzian fluid in Landau frame}

The resulting modes depend strongly on $w^{i}$, the introduced inverse velocity. A similar feature is observed in relation to the dependence of fluid velocity in a Lorentzian fluid in the Landau frame. Let us first look at the case where the fluid is `at rest', namely when $w^i = 0$. We find that there are three diffusive modes. It turns out that the condition $T^i_{\;\;\, 0}=0$ enforces the relation between $\delta T$ and $\delta \mu$ and their fluctuations to decouple from the longitudinal velocity. The dispersion relations for the three diffusive modes are
%
\begin{equation}\label{eq:dispersions}
  \omega_1 = -i \frac{\eta}{\tilde \CE +P} k^2\, ,\quad 
  \omega_2 = -i \frac{\left(\frac{\d \tilde\CE}{\d  T}  \right)_{ \frac{\mu}{T}}\sigma/T  }{\left(\frac{\d \tilde\CE}{\d  T}  \right)_{ \mu}\left(\frac{\d  n}{\d  \mu}  \right)_{ T}- \left(\frac{\d \tilde\CE}{\d  \mu}  \right)_{ T}\left(\frac{\d  n}{\d  T}  \right)_{ \mu}}k^2\, ,\quad 
  \omega_3 = -i  \frac{\zeta + \frac{4}{3}\eta}{\tilde\CE+P} k^2
  \,.
\end{equation}
These modes $\{\omega_1,\omega_2,\omega_3 \}$ correspond to shear diffusion, density diffusion and the longitudinal momentum diffusion respectively. 
The value of $\left(\frac{\d \tilde\CE}{\d  T}  \right)_{ \mu}\left(\frac{\d  n}{\d  \mu}  \right)_{ T}- \left(\frac{\d \tilde\CE}{\d  \mu}  \right)_{ T}\left(\frac{\d  n}{\d  T}  \right)_{ \mu}>0$, since it is nothing but the determinant of the susceptibility matrix that relates fluctuations $( \delta \mu, \delta T)$ to $(\delta \tilde{\mathcal{E}},\delta n)$. The heat capacity in \eqref{eq:dispersions} is supposed to be positive \cite{Kovtun:2012rj}.


 Once the fluid acquires a finite inverse velocity, say $w^i = (w^x,0,0)$, some of these diffusive modes will drive the instabilities, depending on values of thermodynamic quantities, similar to what happens in the Lorentzian fluid. Let us first consider the mode due to the shear diffusion, obtained by solving for $w^{i}$ perpendicular to $k_i$. We find that the spectrum is now described by a quadratic equation in $\omega$ which yields the solutions
\begin{equation}
  \omega^-_1 = -i \frac{\eta}{\tilde \CE+P}  k^2\,, 
  \qquad 
  \omega_1^+ =\frac{2w^i}{ w^2}k_{i}+ i \frac{\tilde \CE+P}{\eta w^2} + i \frac{\eta}{\tilde \CE+P}  k^2\, .
\end{equation}
Hence, the first order Carrollian fluid in the Landau frame is unstable when $w^i \neq0$. One can consider the spectrum of the two other diffusive modes, in the presence of nonzero $w^i$, and find that their spectrum now contains a mode with positive imaginary part, namely
\begin{equation}
  \omega^+_2 = \frac{2 w^i}{w^2}k_{i} + i \frac{\left(\frac{\d \tilde\CE}{\d  T}  \right)_{ \mu}\left(\frac{\d  n}{\d  \mu}  \right)_{ T}- \left(\frac{\d \tilde\CE}{\d  \mu}  \right)_{ T}\left(\frac{\d  n}{\d  T}  \right)_{ \mu}}{\left(\frac{\d \CE}{\d T}\right)_{\frac{\mu}{T}}\sigma w^2/T}  
  + 
  i \frac{\left(\frac{\d \CE}{\d T}\right)_{\frac{\mu}{ T}}\sigma/T 
  }{ 
  \left(\frac{\d \tilde\CE}{\d  T}  \right)_{ \mu}\left(\frac{\d  n}{\d  \mu}  \right)_{ T}- \left(\frac{\d \tilde\CE}{\d  \mu}  \right)_{ T}\left(\frac{\d  n}{\d  T}  \right)_{ \mu} 
  }k^2\,,
\end{equation}
and similarly for the longitudinal momentum diffusion
\begin{equation}
  \omega^+_3 = \frac{2(\zeta+\frac{4}{3}\eta)w^i}{ (\zeta+\frac{4}{3}\eta)w^2} k_{i}+ i \frac{(\tilde \CE + P)}{ (\zeta +\frac{4}{3}\eta)w^2}   + i \frac{\zeta+\frac{4}{3}\eta}{\tilde \CE+P}k^2
  \,.
\end{equation}
Note that all these modes $\omega^+_{1},\omega^+_{2},\omega^+_{3}$ can still cause instabilities even when $k^{i}=0$. Its positive real part is also proportional to $1/w^2$, indicating that the limit where $w^{i}\to 0$ is a singular limit for which the pole $\omega^+ \to +i\infty$ and makes this entire mode decouple. This is the same singular limit observed in Landau frame of Lorentzian fluid in \cite{Hiscock:1985zz}.
\subsection{$c\to 0$ limit of Lorentzian fluid in Eckart frame}
Taking the $c\to 0$ limit of the Lorentzian fluid's Eckart frame puts even stronger constraints on the Carrollian fluid than in the Landau frame. In this case, both $\d_0 \tilde\CE$ and $\d_0 \tilde n$ vanish. This implies that there is no diffusive mode coming from the temperature and chemical potential fluctuations. Nevertheless, the nontrivial linearised dynamics can still be found in the momentum correlators. It turns out, as in the Lorentzian case, that the Eckart frame creates an instability but the limit $w^i \to 0$ is no longer a singular limit. Firstly, the pole in the transverse momentum correlation function is 
\begin{equation}
  \omega_\perp^+ = \frac{2\eta w^i}{\eta w^2 + \kappa T}k^{i} + i \frac{\tilde \CE +P}{\eta w^2 + \kappa T} - \omega^-_\perp,\qquad \omega^-_\perp = -i \frac{\eta}{\tilde\CE + P} k^2\, .
\end{equation}
Similarly, the longitudinal momentum contains two poles. One of them describes momentum diffusion, while the other one causes the instability 
\begin{equation}
\begin{aligned}
  \omega_\parallel^+ &= \frac{2(\zeta+2\eta)k_i w^i}{(\zeta +\frac{4}{3}\eta)w^2 + \kappa T} + i \frac{\tilde\CE+P}{(\zeta+\frac{4}{3}\eta)w^2 + \kappa T}  - \omega^-_\parallel\,,\\
  \omega_\parallel^- &= -i \frac{\zeta + \frac{4}{3}\eta}{\tilde\CE + P} k^2
  \,.
\end{aligned}
\end{equation}
Hence, the Carrollian fluid in the Eckart frame is also unstable, even when $P_i \sim w^i =0$. We can also see that, not only the Eckart frame signals the instability in the $w^i = 0$ limit. This frame choice also changes the number of diffusive poles in the spectrum, which are supposed to be physical objects. The origin of this mismatch might be the fact that from a Carroll point of view, the Eckart and Landau frame seem not to be connected by a suitable redefinition of parameters, as discussed in Section \ref{sec:car_intro}. The definitions for what is Eckart and Landau frame in this context is inherited from tracking the Lorentzian case.
\subsection{$c\to 0$ limit of Lorentzian fluid in general frame}
In the general frame, we treat $\kappa$ and $\sigma$ as general variables. The general modes we obtain, are the density diffusion (coming from the fluctuations in chemical potential or temperature)
\begin{equation}
  \omega^+_{1} 
 \! =\!
  \frac{2 w^i}{w^2}k^{i} 
  + 
  i \frac{ 
  \left(\frac{\d \tilde\CE}{\d  T}  \right)_{ \mu}
  \!\!
  \left(\frac{\d  n}{\d  \mu}  \right)_{ T}
  - 
  \left(\frac{\d \tilde\CE}{\d  \mu}  \right)_{ T}
  \!\!
  \left(\frac{\d  n}{\d  T}  \right)_{ \mu} 
  }{
  \left(\frac{\d \CE}{\d T}\right)_{\frac{\mu}{T}}\sigma w^2/T}  
-
\omega^{-}_{1}\!\!\,,
  \quad
  \omega^{-}_{1}
  \!=\!
-i \frac{\left(\frac{\d \tilde\CE}{\d  T}  \right)_{ \frac{\mu}{T}}\sigma/T
}{
\left(\frac{\d \tilde\CE}{\d  T}  \right)_{ \mu}
\!\!
\left(\frac{\d  n}{\d  \mu}  \right)_{ T}- \left(\frac{\d \tilde\CE}{\d  \mu}  \right)_{ T}
\!\!
\left(\frac{\d  n}{\d  T}  \right)_{ \mu} 
}k^2
\!
\,.
\end{equation}
The following modes are associated to longitudinal momentum
\begin{equation}
  \omega_\parallel^+ = \frac{2(\zeta+\frac{4}{3}\eta) w^i}{w^2(\zeta +\frac{4}{3}\eta) + \kappa T}k^{i} + i \frac{\tilde\CE+P}{w^2(\zeta+\frac{4}{3}\eta) + \kappa T}  - \omega^-_\parallel\,,
  \quad
  \omega_\parallel^- = -i \frac{\zeta + \frac{4}{3}\eta}{\tilde\CE + P} k^2
  \,.
\end{equation}
Finally, the following modes are coming from the transversal channel (with multiplicity $d-1$)
\begin{equation}
  \omega_\perp^+ = \frac{2 \eta w^i}{\eta w^2 + \kappa T}k^{i} + i \frac{\tilde \CE +P}{\eta w^2 + \kappa T} - \omega^-_\perp,\qquad \omega^-_\perp = -i \frac{\eta}{\tilde\CE + P} k^2\, .
\end{equation}
Putting $\sigma=0$ or $\kappa=0$, successfully reproduces the Eckart frame and Landau frame, respectively. Only in the Landau frame, one is able to find stability when $P_i \propto w^i = 0$.

\section{Stability beyond boost symmetry}\label{generalfluid}
The goal of this section is to research instabilities away from boost symmetries, depending on frame choice. We will analyse this using the linearized static fluid introduced in Section \ref{sec:general}.

Combing the input of the Landau frame \eqref{eq:genlau} and Eckart frame \eqref{eq:geneck}, we can compute eigenmodes for the linearised spectrum in momentum space. If we consider $\delta u^{i}=(u,0,..,0)$ and $k_{i}=(0,k,...,0)$, we can isolate the following two modes
\begin{equation}\label{eq:gen_modes}
	\omega_{+}
	=
	i\frac{\rho}{A}
	+
	i\frac{\eta}{\rho}k^{2}
	\,,
	\quad
	\omega_{-}
	=
	-
	i
	\frac{\eta}{\rho}k^{2}
	\,,
\end{equation}
where 
\begin{equation}
    A = \left\{\begin{array}{lr}
        \overline{\pi}-\tilde{a}_{T}, &\quad\text{for Landau frame}\\
        \overline{\pi}-\tilde{a}_{T}+\left(\frac{\rho}{n}+1\right)(\overline{\alpha}-\tilde{a}_{\frac{\mu}{T}})+\frac{\rho}{n}\sigma, &\quad\text{for Eckart frame.}
        \end{array}\right.
\end{equation}
Since $\text{Im}(\omega_{+}) > 0$ in \eqref{eq:gen_modes}, there will always be an instability (for small enough momentum $k$), unless $A\leq0$. It turns out that it is therefore, a priori, unclear whether a fluid, without boost perse, is unstable or not.
 Let us now examine what happens in various boost invariant scenarios.
\\
\\
\textbf{Lorentz.} For the Lorentzian case we have $\overline{\pi}=\overline{\alpha}=\tilde{a}_{T}=\tilde{a}_{\frac{\mu}{T}}=0$ and $\rho=\tilde{\mathcal{E}}+P$, due to the Lorentz Ward identity $T^{i}_{\;\;0}=T^{0}_{\;\;i}$. It is immediate that the instability is absent (at least for a static fluid) from the spectrum in Landau frame, since $A=0$. However, in the Eckart frame we find that the coefficient becomes
\begin{equation}
			A
			=
				\overline{\pi}
			+
			\left(
			\frac{\rho}{n}+1
			\right)\overline{\alpha}
			+
			\frac{\rho}{n}\sigma
			=
			\frac{\tilde{\mathcal{E}}+P}{n}\sigma
			\,,
\end{equation}
which in principle is strictly positive. As a result, the instability persists in the Eckart frame. By reinserting $c$ and subsequently taking $c\to\infty$, one is expected to retain the same conclusions for \textbf{Carroll}. 
\\
\\
\textbf{Bargmann.} For the Bargmann case $\overline{\pi}=-\overline{\alpha}=\sigma$, $\tilde{a}_{T}=\tilde{a}_{\frac{\mu}{T}}=0$ and $\rho=n$, due to the Bargmann Ward identity $T^{0}_{\;\;i}=mJ^{i}$. It  is immediately clear that the Landau frame remains unstable. For the Eckart frame, we however find that the coefficient 
\begin{equation}
			A
			=
					\overline{\pi}
			+
			\left(
			\frac{\rho}{n}+1
			\right)\overline{\alpha}
			+
			\frac{\rho}{n}\sigma
			=
			\overline{\alpha}
			+
			\sigma
			=
			0
			\,,			
\end{equation}
which implies that the instability, for a static fluid, is absent in the Eckart frame.

\newpage
\section*{Acknowledgements}
We would like to thank S. Grozdanov, J. Hartong, P. Kovtun, A. Lucas, N. Obers, G. Oling, and L. Thorlacius for discussions. We would also like to thank S. Grozdanov, J. Hartong and P. Kovtun for commenting on the manuscript. The work of N. P. was supported by Icelandic Research Fund (IRF) grant 163422-052. The research of W. S. is funded by IRF Postdoctoral Fellowship Grant 185371-051. The authors would like to thank NORDITA, the Niels Bohr Institute, the Max Planck Institute for physics of complex systems and Chulalongkorn University for their hospitality.

\begin{appendix}
\section{Changing frames}\label{appendix1}

The goal of this appendix is to derive how we can switch between frames for the fluid considered in Section \ref{sec:general}. Specifically, we start from the constitutive relations given in \eqref{eq:aristotle_const} with $\tilde{E}_{2}=0$, $\mathcal{P}_{2}=0$, $\mathcal{M}_{2}=0$, $\mathcal{N}_{2}=0$, $\overline{q}_{2\nu}=0$. The result of this appendix is equation \eqref{eq:relating}.

We will consider a general redefinition of the form
\begin{equation}
	T
	\to
	T
	+
	\eth T
	\,,\quad
	\mu
	\to
	\mu
	+
	\eth \mu
	\,, \quad
	u^{\mu}
	\to
	u^{\mu}
	+
	\eth u^{\mu}
	\,,
\end{equation}
where all terms with an $\eth$ are of order $\mathcal{O}(\partial)$. If we apply such a change of variables, we obtain, keeping up to $\mathcal{O}(\partial)$ terms,
\begin{equation}\begin{aligned}
	T^{\mu}_{\;\;\,\nu}(T+\eth T,\mu+\eth\mu,u^{\mu}+\eth u^{\mu})
	=&\;
	T^{\mu}_{\;\;\,\nu}(T,\mu,u^{\mu})
	+
	 u^{\mu}\overline{u}_{\nu}\eth \tilde{\mathcal{E}}
	+
	\Delta^{\mu}_{\;\;\,\nu}\eth P
	+
	\left(\tilde{\mathcal{E}} \overline{u}_{\nu}
		+
		P\overline{u}_{\nu}
		+
		\rho \Gamma_{\nu}
	\right)\eth u^{\mu}
	\\
	&
	+
	\overline{u}_{\nu}\eth q^{\mu}
	+
	u^{\mu}\Gamma_{\nu}\eth \rho
	+
	\rho u^{\mu}\eth \Gamma_{\nu}
	+
	u^{\mu}\eth \overline{q}_{\nu}
	+
	\eth t^{\mu}_{\;\;\,\nu}
	+
	\mathcal{O}(\partial^{2})
	\,,
	\\
	J^{\mu}(T+\eth T,\mu+\eth\mu,u^{\mu}+\eth u^{\mu})
	=&\;
	J^{\mu}(T,\mu,u^{\mu})
	+
	u^{\mu}\eth n
	+
	\eth u^{\mu}
	+
	\eth j^{\mu}
	+
	\mathcal{O}(\partial^{2})
	\,,
\end{aligned}\end{equation}
where we dropped all subscripts in order avoid cluttered notation.

Again, terms with an $\eth$ are of order $\mathcal{O}(\partial)$ and share the same symmetry properties as their `parent'. Because we will be interested in linearisation around a zero velocity background, we can put $i^{\mu}$ and $\eth i^{\mu}$ to zero without any loss of generality, since these terms are always accompanied by $u_{\nu}$.

From thermodynamical considerations we can express
\begin{equation}\begin{aligned}
	\eth
	\mathcal{I}
	=&
	\left(
		\frac{\partial \mathcal{I}
		}{
		\partial T}
	\right)
	\eth T
	+
	\left(
		\frac{\partial \mathcal{I}
		}{
		\partial \mu}
	\right)
	\eth \mu
	+
	\left(
		\frac{\partial \mathcal{I}
		}{
		\partial u^{\mu}}
	\right)
	\eth u^{\mu}
	\,,
\end{aligned}\end{equation}
where $\mathcal{I}\in\left\{\tilde{\mathcal{E}},n,P,\rho\right\}$.
We will use velocity to switch frames, but $\eth\mu$ and $\eth T$ are still free to choose. We make the usual choice, to pick $\eth \mu$ and $\eth T$ in such a way that $\eth \tilde{\mathcal{E}}=0$ and $\eth n=0$. Notice that in the Bargmann case, $m n=\rho$, where $m$ is particle mass, such that $\delta\rho=0$ too. 

In order to be able to make a choice of frame, we express the changes in $\eth j^{\alpha}$, $\eth P$, $\eth t^{\mu}_{\;\;\,\nu}$, $\eth q^{\mu}$ and $\eth \overline{q}_{\nu}$ in terms of $\eth u^{\mu}$ by requiring that $T^{\mu}_{\;\;\,\nu}(T+\eth T,\mu+\eth\mu,u^{\mu}+\eth u^{\mu})=T^{\mu}_{\;\;\,\nu}(T,\mu,u^{\mu})$ and $J^{\mu}_{\;\;\,\nu}(T+\eth T,\mu+\eth\mu,u^{\mu}+\eth u^{\mu})=J^{\mu}_{\;\;\,\nu}(T,\mu,u^{\mu})$. 

Let us consider the constitutive relations linearised around a stationary background (i.e. $u^{\mu}=\tau^{\mu}+\delta u^{\mu}$), it happens that the definition for Landau and Eckart frame for Bargmann and Lorentz coincide. The Carroll case follows from the Lorentzian one. We aim to eliminate $\eth u^{\mu}$ in terms of other `$\eth$' objects. In order to do that, we first establish
\begin{equation}\begin{aligned}
	j^{\alpha}
	:=
	\Delta^{\alpha}_{\;\;\,\mu}J^{\mu}
	=
	j^{\alpha}
	+
	n\eth u^{\alpha}
	+
	\eth j^{\alpha}
	\,,
	\quad
	\Rightarrow
	&
	\quad
	\eth j^{\alpha}
	=
	-n\eth u^{\alpha}
	\,,
	\\
	\mathcal{P}
	:=
	\frac{1}{d}
	\Delta^{\nu}_{\;\;\mu}T^{\mu}_{\;\;\,\nu}
	=
	\mathcal{P}
	+
	\eth P
	+
	\frac{1}{d}\rho \Gamma_{\nu}\eth u^{\nu}
	\,,
	\quad
	\Rightarrow
	&
	\quad
	\eth P
	=
	-\frac{1}{d}\rho \Gamma_{\nu}\eth u^{\nu}
	\,,
	\\
	t^{\mu}_{\;\;\,\nu}
	:=
	\Delta^{\mu}_{\;\;\rho}\Delta^{\sigma}_{\;\;\nu}T^{\rho}_{\;\;\sigma}
	-
	\frac{1}{d}\Delta^{\rho}_{\;\;\sigma}T^{\sigma}_{\;\;\rho}\Delta^{\mu}_{\;\;\nu}
	\qquad\qquad\qquad\quad\;\;\;
	&
	\\
	=
	t^{\mu}_{\;\;\nu}
	+
	\eth t^{\mu}_{\;\;\nu}
	+
	\rho \Delta^{\lambda}_{\;\;\nu}\Gamma_{\lambda}\eth u^{\mu}
	-
	\rho \Gamma_{\lambda}\eth u^{\lambda}\Delta^{\mu}_{\;\;\,\nu}
	\,,
	\quad
	\Rightarrow
	&
	\quad
	\eth t^{\mu}_{\;\;\nu}
	=
	\rho \Gamma_{\lambda}\eth u^{\lambda}\Delta^{\mu}_{\;\;\,\nu}
	-
	\rho \Delta^{\lambda}_{\;\;\nu}\Gamma_{\lambda}\eth u^{\mu}
	\,,
	\\
	q^{\mu}
	:=
	-
	\Delta^{\mu}_{\;\;\rho}u^{\sigma}T^{\rho}_{\;\;\,\sigma}
	=
	q^{\mu}
	+
	\eth q^{\mu}
	+
	\tilde{\mathcal{E}}\eth u^{\mu}
	+
	P \eth u^{\mu}	
	\,,
	\quad
	\Rightarrow
	&
	\quad
	\eth q^{\mu}
	=
	-
	(
	\tilde{\mathcal{E}}
	+
	P) \eth u^{\mu}	
	\,,
	\\
	\overline{q}_{\nu}
	+
	\mathcal{M}
	\Gamma_{\nu}
	:=
	-\Delta^{\sigma}_{\;\;\nu}\overline{u}_{\rho}T^{\rho}_{\;\;\sigma}\,\qquad\qquad\qquad\qquad\qquad\qquad&
	\\
	=
	\overline{q}_{\nu}
	+
	\mathcal{M}
	\Gamma_{\nu}
	+
	\Gamma_{\nu}
	\eth \rho
	+
	\rho \Delta^{\alpha}_{\;\;\,\nu}
	\eth\Gamma_{\alpha}
	+
	\eth \overline{q}_{\nu}
		\Rightarrow&
	\quad
	\eth \overline{q}_{\nu}
	+
	\eth \rho
	\Gamma_{\nu}
	=
	-
	\rho
	\Delta^{i}_{\;\;\,\nu}\eth u^{i}
	\,,
\end{aligned}\end{equation}
where in the last line we have made use of the identity $\eth \Gamma_{\alpha}= \overline{u}_{\alpha}u^{i}\eth u^{i}+\Delta^{\beta}_{\;\;\,\alpha}\Delta^{i}_{\;\;\,\beta}\eth u^{i}$. 

A new frame (denoted by tilde) can thus be related to an old frame (no tilde) via
\begin{equation}
	\tilde{E}'=\tilde{E}
	=
	\tilde{\mathcal{E}}
	\,,
	\quad
	\mathcal{N}'
	=
	\mathcal{N}
	=
	n
	\,,
	\quad
	\mathcal{P}'
	=
	\mathcal{P}
	-
	\frac{1}{d}\rho\Gamma_{\nu}\eth u^{\nu}
	\,,
	\quad
	\overline{q}'_{\nu}
	+
	\mathcal{M}'
	\Gamma_{\nu}
	=
	\overline{q}_{\nu}
	+
	\mathcal{M}
	\Gamma_{\nu}
	-
	\rho
	\Delta^{i}_{\;\;\,\nu}\eth u^{i}
	\,,
\end{equation}
\begin{equation}
	j'^{\mu}
	=
	j^{\mu}
	-
	n \eth u^{\mu}
	\,,
	\quad
	q'^{\mu}
	=
	q^{\mu}
	-
	(\tilde{\mathcal{E}}+P)\eth u^{\mu}
	\,,
	\quad
	t'^{\mu}_{\;\;\,\nu}
	=
	t^{\mu}_{\;\;\,\nu}
	+
	\rho \Gamma_{\lambda}\eth u^{\lambda}\Delta^{\mu}_{\;\;\,\nu}
	-
	\rho \Delta^{\lambda}_{\;\;\nu}\Gamma_{\lambda}\eth u^{\mu}
	\,.
\end{equation}
By choosing $\delta u^{\mu}=\frac{1}{n}j^{\mu}$, we can use the equations above to relate the Eckart frame (denoted by subscript $E$) and Landau frame (denotes by subscript $L$):
\begin{equation}\label{eq:switch1}
	E_{E}=E_{L}=\tilde{\mathcal{E}}
	\,,
	\quad
	\mathcal{N}_{E}
	=
	\mathcal{N}_{L}
	=
	n
	\,,
	\quad
	\mathcal{P}_{E}
	=
	\mathcal{P}_{L}
	-
	\frac{1}{d}\frac{\rho}{n}\Gamma_{\nu} j^{\nu}_{L}
	\,,
	\quad
	\overline{q}_{E\,\nu}
	+
	\mathcal{M}_{E}
	\Gamma_{\nu}
	=
	\overline{q}_{L\,\nu}
	+
	\mathcal{M}_{L}
	\Gamma_{\nu}
	-
	\frac{\rho}{n}
	\Delta^{i}_{\;\;\,\nu} j^{i}_{L}
	\,,
\end{equation}
\begin{equation}\label{eq:switch2}
	j_{E}^{\mu}
	=
	0
	\,,
	\quad
	q^{\mu}_{E}
	=
	q^{\mu}_{L}
	-
	\frac{\tilde{\mathcal{E}}+P}{n} j_{L}^{\mu}
	\,,
	\quad
	t^{\mu}_{E\;\nu}
	=
	t^{\mu}_{L\;\nu}
	+
	\frac{1}{d}
	\frac{\rho}{n} \Gamma_{\lambda}j^{\lambda}\Delta^{\mu}_{\;\;\,\nu}
	-
	\frac{\rho}{n} \Delta^{\lambda}_{\;\;\nu}\Gamma_{\lambda}j^{\mu}
	\,.
\end{equation}
We finally obtain the relations between Landau and Eckart frame
\begin{equation}\label{eq:relating}
	 \mathcal{I}_{E}
	=
	 \mathcal{I}_{L}
	\,,
	\quad	
	\delta \overline{q}_{E\,\nu}
	=
	\delta \overline{q}_{L\,\nu}
	-
	\frac{\rho}{n}
	h_{\nu\mu}\delta j^{\mu}_{L}
	\,,
	\quad
	\delta q^{\mu}_{E}
	=
	-
	\frac{\tilde{\mathcal{E}}+P}{n} \delta j_{L}^{\mu}
	\,,
\end{equation}
where $\delta$ denotes a linearised fluctuation, the capital letters as subscript denote respective frame and $\mathcal{I}\in\left\{\tilde{\mathcal{E}},P,n,\rho,\delta\tilde{E},\delta\mathcal{N},\delta\mathcal{P},\delta t^{\mu}_{\;\;\nu}\right\}$. The Eckart frame implies $\delta j^{\mu}_{E}=0$ and the Landau frame implies $\delta q^{\mu}_{L}=0$.

\end{appendix}

 \bibliographystyle{utphys}
\bibliography{ref.bib}

\end{document}